\documentclass[11pt,a4paper]{article}
\usepackage{jheppub}

\usepackage{axodraw}

\newcommand{\pt}{p_{_T}}
\newcommand{\xt}{x_{_T}}
\def\cO#1{{{\cal{O}}}\left(#1\right)}
\newcommand{\sqrtsnn}{\sqrt{s_{_{\rm NN}}}}

\def\msbar{\ensuremath{{\rm{\overline{MS}}}}}

\newcommand{\mui}{\ensuremath{\mu_F}}

\newcommand{\mur}{\ensuremath{\mu_R}}

\title{Probing gluon and heavy-quark nuclear PDFs with $\gamma+Q$ production in $pA$ collisions}

\author[a]{T.\ Stavreva,}
\author[a]{I.\ Schienbein,}
\author[b]{F.\ Arleo,}
\author[c]{K.\ Kova\v{r}\'{\i}k,}
\author[d]{F.\ Olness,}
\author[d]{J.\ Y.\ Yu}
\author[e]{and \\ 
J.\ \ F.\ Owens}

\affiliation[a]{Laboratoire de Physique Subatomique et de Cosmologie, UJF, CNRS/IN2P3,
INPG, \\
53 avenue des Martyrs, 38026 Grenoble, France} 
\affiliation[b]{Laboratoire d'Annecy-le-Vieux de Physique Th\'eorique (LAPTH) 
UMR5108, Universit\'e de Savoie, CNRS,\\
BP 110, 74941 Annecy-le-Vieux cedex, France}
\affiliation[c]{Karlsruhe Institute of Technology (KIT), Fakult\"at f\"ur Physik, 
Institut f\"ur Theoretische Physik (IThP),\\
Postfach 6980, D-76128 Karlsruhe}
\affiliation[d]{Southern Methodist University,\\
Dallas, TX 75275, USA}
\affiliation[e]{Florida State University,\\
 Tallahassee, FL 32306, USA}

\emailAdd{stavreva@lpsc.in2p3.fr}
\emailAdd{schien@lpsc.in2p3.fr}
\emailAdd{arleo@lapp.in2p3.fr}
\emailAdd{kovarik@particle.uni-karlsruhe.de}
\emailAdd{olness@physics.smu.edu}
\emailAdd{yu@physics.smu.edu}
\emailAdd{owens@hep.fsu.edu}

\abstract{
We present a detailed phenomenological study of direct photon production in
association with a heavy-quark jet in $pA$ collisions at the Relativistic Heavy Ion Collider (RHIC) 
and at the Large Hadron Collider (LHC) at next-to-leading order in QCD. 
The dominant contribution to the cross-section comes from the gluon--heavy-quark
($gQ$) initiated subprocess, making $\gamma + Q$ production a process very sensitive to both the gluon 
and the heavy-quark parton distribution functions (PDFs). 
 Additionally, the RHIC and LHC experiments are probing complementary kinematic regions in the 
momentum fraction $x_{_2}$ carried by the target partons.
Thus, the nuclear production ratio $R^{\gamma+Q}_{pA}$ can provide strong constraints, over a broad $x$-range, on the poorly determined nuclear parton distribution functions which are extremely important for the interpretation of 
results in heavy-ion collisions.

}


\begin{document} 

\maketitle

\section{Introduction}
\label{sec:intro}

\noindent 
Parton distribution functions (PDFs) are an essential component of any prediction involving colliding hadrons. 
The PDFs are non-perturbative objects which have to be determined from experimental input and link theoretical 
perturbative QCD (pQCD) predictions to observable phenomena at hadron colliders. 
In view of their importance, the proton PDFs have been a focus of long and dedicated global analyses performed 
by various groups; see e.g. Refs.~\cite{Nadolsky:2008zw,Martin:2009iq,Ball:2009mk,Ball:2008by,JimenezDelgado:2008hf,JimenezDelgado:2009tv,Alekhin:2010iu,Alekhin:2009ni} for some of the most recent studies.
Over the last decade, global analyses of PDFs in nuclei --~or nuclear PDFs (nPDFs)~-- have been performed by several groups:
nCTEQ~\cite{Schienbein:2007fs,Schienbein:2009kk,Kovarik:2010uv}, nDS~\cite{deFlorian:2003qf}, EKS98~\cite{Eskola:1998df}, EPS08/EPS09~\cite{Eskola:2008ca,Eskola:2009uj},  and  HKM/HKN~\cite{Hirai:2001np,Hirai:2004wq,Hirai:2007sx} (for a recent review, see Ref.~\cite{Armesto:2006ph}).
In a manner analogous to the proton PDFs, the nPDFs are needed in order to predict observables in
proton--nucleus ($pA$) and nucleus--nucleus ($AA$) collisions. 
However, as compared to the proton case, the nuclear parton distribution functions are far less well constrained.
Data that can be used in a global analysis are available for fewer hard processes and also cover a smaller 
kinematic range.
In particular, the nuclear gluon distribution is only very weakly constrained, leading to a significant uncertainty in the theoretical predictions of hard processes in $AA$ collisions.

%
For this reason it is crucial to use a variety of hard processes in $pA$ collisions, both at RHIC and at LHC,
in order to better constrain nuclear parton densities. 
The inclusive production of jets, lepton pairs or vector bosons 
are natural candidates since they are already used in global analyses of proton PDFs\footnote{Recently, 
a paper by Paukkunen and Salgado~\cite{Paukkunen:2010qg} discussed that weak boson production at the LHC 
might be useful in order to constrain nPDFs.}.
In addition, other processes which could constrain the gluon nPDF have been discussed in the literature and have yet to be employed.
For instance, the production of isolated direct photons~\cite{Arleo:2007js} as well as inclusive 
hadrons~\cite{QuirogaArias:2010wh} at RHIC and LHC can provide useful constraints on the nuclear 
gluon distribution\footnote{Note that the EPS08/EPS09~\cite{Eskola:2008ca,Eskola:2009uj} global analyses 
include single-inclusive $\pi^0$ data from the PHENIX experiment at RHIC.}, even though in the latter 
channel the fragmentation process 
complicates its extraction.
Another natural candidate for measuring the gluon nPDF is heavy-quark~\cite{Eskola:2001gt} 
or heavy-quarkonium~\cite{BrennerMariotto:2009ey} production. 
Quarkonium production is however still not fully under control theoretically (see e.g.~\cite{Lansberg:2006dh} 
for a review), hence it is not obvious whether a meaningful extraction of the nuclear gluon PDF will 
eventually be possible in this channel, yet indirect constraints might be obtained~\cite{Arleo:2008zc}.
%

In this paper, we investigate the production of a direct photon in association
with a heavy-quark jet in $pA$ collisions
in order to constrain parton densities in nuclei{\footnote {A recent paper by Betemps and Machado~\cite{Betemps:2010ay} has performed a calculation for $\gamma + c$ production using however the target rest frame formalism.}}.
As we will show, this process is dominated by the 
heavy-quark--gluon ($Qg$) initial state at both RHIC and the LHC
making the nuclear production ratio in $pA$ over $pp$ collisions,
\begin{equation}
R_{pA}^{\gamma Q}= 
\frac{\sigma\left(pA\to\gamma\ Q\ {\rm X}\right)}{A\ \sigma\left(pp\to\gamma\ Q\ {\rm X}\right)}\, ,
\label{eq:ratio1}
\end{equation}
a useful observable in order to determine the gluon and heavy-quark nPDFs in 
complementary $x$-ranges from RHIC to LHC.
One of the advantages of such a ratio is that many of the experimental and 
theoretical uncertainties cancel.
Nevertheless, for a solid interpretation of the ratios it is also necessary to compare
the theory directly with the (differential) measured cross-sections.
For this reason we present cross-sections and $p_T$-distributions computed at next-to-leading order (NLO)
of QCD using acceptance and isolation cuts appropriate for the PHENIX and ALICE experiments
at RHIC and LHC, respectively.
Using the available luminosity values we also provide simple estimates for the expected
event numbers.

The paper is organized as follows. 
In section~\ref{sec:framework} we briefly describe the NLO calculation used in the present 
paper (more details can be found in~\cite{Stavreva:2009vi,Stavreva:PhD}).
In section \ref{sec:PDF}, we discuss the different nPDF sets used in our analysis, focusing 
especially on the gluon and the heavy-quark sectors.
In  sections~\ref{sec:pA_RHIC} and~\ref{sec:pA_LHC}, results in $pA$ collisions at RHIC and LHC, respectively,
are presented.
In each case, we start with a discussion of the acceptance and isolation cuts, then turn to the 
(differential) cross-sections and event numbers, followed by a discussion of the nuclear production
ratios.
Finally, we summarize our main results in section \ref{sec:conclusions}.
%



\section{Direct photon production in association with a heavy-quark jet}
\label{sec:framework}

Single direct photons have long been considered an excellent probe of the structure of the proton due to 
their point-like electromagnetic coupling to quarks and due to the fact that they escape 
confinement~\cite{Aurenche:1988vi,Owens:Rev}.  
Their study can naturally be extended to high-energy nuclear collisions where one can use 
direct photons to investigate the structure of nuclei as well ~\cite{Arleo:2007js}.

However, it might also be
relevant to study more exclusive final states, such as the double inclusive production of a direct photon
in association with a heavy-quark (charm, bottom) jet~\footnote{We do not distinguish between heavy-quark 
and heavy-anti-quark in the final state, that is the sum of $\gamma+Q$ and $\gamma + \bar Q$ is considered; therefore here by $qQ\rightarrow qQ \gamma$ we choose to denote the sum of $qQ\rightarrow qQ \gamma$, 
$q\bar Q\rightarrow q \bar Q \gamma$, $\bar qQ\rightarrow \bar qQ \gamma$ and $\bar q \bar Q\rightarrow \bar q \bar Q \gamma$.} in order to get additional  constraints on parton distribution functions. The lower counting rates expected for this observable are compensated by various advantages:
\begin{itemize}
\item As shown below, the cross-section for direct photon plus heavy-quark production
in $pp$ and $pA$ collisions is largely dominated by the gluon--heavy-quark ($gQ$) channel. 
This offers in principle a direct access to the gluon and heavy-quark distributions in a proton and in nuclei;
\item A two-particle final-state allows for the independent determination of the parton momentum fractions
$x_1$ (projectile) and $x_2$ (target), using leading order kinematics and in the absence of 
fragmentation processes;
\item Since the valence up quark distribution (to which single photons mostly couple) is smaller in 
neutrons --~and therefore in nuclei~-- as compared to that in a proton, 
the nuclear production ratio $R_{pA}^\gamma$ of {\it single} photon production at large $\xt=2\pt/\sqrt{s}$ 
is different than 1 independently of any nPDF effects~\cite{Arleo:2006xb}.
In the $\gamma+Q$ production channel the photon couples mostly to the heavy-quark, which, by isospin symmetry,
has the same distribution in a proton or neutron, i.e. $Q^p=Q^n$, leading to a nuclear 
production ratio $R_{pA}^{\gamma Q}$ free of any ``isospin'' effects and thus properly normalized to 1 
in the absence of nPDF corrections.
\end{itemize}

At leading-order accuracy, ${\cal O}(\alpha\alpha_s)$, at the hard-scattering level the production of a {\it direct} photon with a heavy-quark jet only arises from the $g Q \to \gamma Q$ Compton scattering process, making this observable highly sensitive to both the gluon and heavy-quark PDFs. This is at variance with the single photon channel for which the Compton scattering ($gq \to \gamma q$) as well as annihilation process ($q \bar q \rightarrow \gamma g$) channels compete\footnote{In $pp$ and $pA$ collisions, however, the Compton subprocess largely dominates the annihilation process from the dominance of the gluon distribution over that of sea-quarks.}.
At NLO the number of contributing subprocesses increases to seven, listed in Table~\ref{table:NLO}.  
As can be seen, all subprocesses apart from $q\bar q\rightarrow \gamma Q\bar Q$ are $g$ and/or $Q$ initiated.  
Which of these subprocesses dominate is highly dependent on the collider type ($p\bar{p}$ {\it vs.} $pp$/$pA$) 
and the collider center-of-mass energy.  
For example, $g$ and $Q$ initiated subprocesses will be more dominant at $pp$ and $pA$ colliders, 
whereas at the Tevatron ($p\bar{p}$ collisions) the $q\bar q\rightarrow \gamma Q\bar Q$  dominates at high $\pt$ 
because of the valence--valence $q\bar{q}$ scattering in these collisions.

\begin{table}[h]
\begin{center}
\begin{tabular}{cc}
\hline
\hline
$gg\rightarrow \gamma Q\bar Q$ &
$gQ\rightarrow \gamma gQ$\\
$Qq\rightarrow \gamma qQ$ &
$Q\bar q\rightarrow \gamma \bar qQ$ \\
$Q\bar Q\rightarrow \gamma Q\bar Q$&
$QQ\rightarrow \gamma QQ$ \\
$q\bar q\rightarrow \gamma Q\bar Q$\\
\hline
\hline
\end{tabular}
\caption{\label{table:NLO}List of all $2\rightarrow 3$ NLO hard-scattering 
subprocesses.}
\end{center}
\end{table}

When one considers higher order subprocesses, such as $qQ \rightarrow qQ\gamma $, the produced photon may be emitted collinearly with the final state $q$ giving rise to a collinear singularity. %
This singular contribution is absorbed in fragmentation functions (FFs) 
$D_{\gamma/q}(z,\mu^2)$, which satisfy a set of inhomogeneous DGLAP equations, the solutions 
of which are of order $\cO{\alpha/\alpha_s}$.  
As a consequence, another class of contributions of order $\cO{\alpha \alpha_s}$ consists of 
$2\rightarrow 2$ QCD subprocesses with at least one heavy-quark in the final state 
and another parton fragmenting into a collinear photon. 
These so-called fragmentation contributions need to be taken into account at each order in the 
perturbative expansion.  
As in the LO direct channel, we also include the $\cO{\alpha\alpha_s^2}$ fragmentation contributions, which 
are needed for a complete NLO calculation.  
It should however be mentioned that isolation requirements --~used experimentally in order to 
minimize background coming from hadron decays~-- greatly decrease these fragmentation contributions.    

The present calculations have been carried out using the strong coupling constant 
corresponding to the chosen PDF set: 
$\alpha_s^{\msbar,5}(M_Z)=0.118$ in next-to-leading order for both nCTEQ and EPS09, and 
$\alpha_s^{\msbar,5}(M_Z)=0.1165$ for HKN. The renormalization, factorization and fragmentation scales have been set to 
$\mur=\mui=\mu_f=p_{T\gamma}$ and we have used $m_c=1.3$ GeV and $m_b=4.5$ GeV for the charm 
and bottom quark masses. We utilize the photon fragmentation functions of L. Bourhis, M. Fontannaz and 
J.P. Guillet \cite{Bourhis:1997yu}. For further details on the theoretical calculations, the 
reader may refer to~\cite{Stavreva:2009vi,Stavreva:PhD}. 


\section{Nuclear Parton Distribution Functions}
\label{sec:PDF}

In order to obtain results in hadronic collisions, the partonic cross-sections
have to be convoluted with PDFs for protons and nuclei.
For the latter we show results using the most recent 
nCTEQ~\cite{Schienbein:2007fs,Schienbein:2009kk}, EPS09~\cite{Eskola:2009uj}, 
and HKN07~\cite{Hirai:2007sx} nuclear PDF sets\footnote{Note that
the nDS04 PDFs~\cite{deFlorian:2003qf} are not considered here since these are obtained in a 
3-fixed flavor number scheme (no charm PDF) whereas our calculation is in a 
variable flavor number scheme.}.
Each set of nuclear PDFs is connected to a set of proton PDFs to which it reduces in the limit $A \to 1$
where $A$ is the atomic mass number of the nucleus\footnote{More precisely, EPS09 is linked to the 
CTEQ6.1M proton PDFs~\cite{Stump:2003yu},
HKN07 to the MRST98~\cite{Martin:1998sq} set, 
and the nCTEQ PDFs to the reference PDFs described in Ref.~\cite{Owens:2007kp}
which are very similar to the CTEQ6.1M distribution functions~\cite{Stump:2003yu}. 
This reference set excludes most of the nuclear data used in the PDF global fit, and therefore is not 
biased by any nuclear corrections.}.  Therefore we use the various nPDFs together with their corresponding 
proton PDFs in the calculations.
Since our goal is to probe gluon and heavy-quark nPDFs, let us now discuss these specific 
distributions in greater detail.

\subsection{Gluon sector}
As already mentioned, the nuclear gluon distribution is only very weakly
constrained in the $x$-range $0.02\lesssim x \lesssim 0.2$ from the $Q^2$-dependence 
of structure function ratios in deep-inelastic scattering (DIS) \cite{Gousset:1996xt}, 
$F_2^{Sn}(x,Q^2)/F_2^{C}(x,Q^2)$, measured by the NMC collaboration~\cite{Arneodo:1996rv}\footnote{As 
discussed earlier, EPS08/EPS09 also include inclusive $\pi^0$ data from the PHENIX experiment at RHIC, with 
a strong weight in order to better determine the nuclear gluon distribution.}.

\begin{figure}
\begin{center}
\includegraphics[angle=-90,scale=0.26]{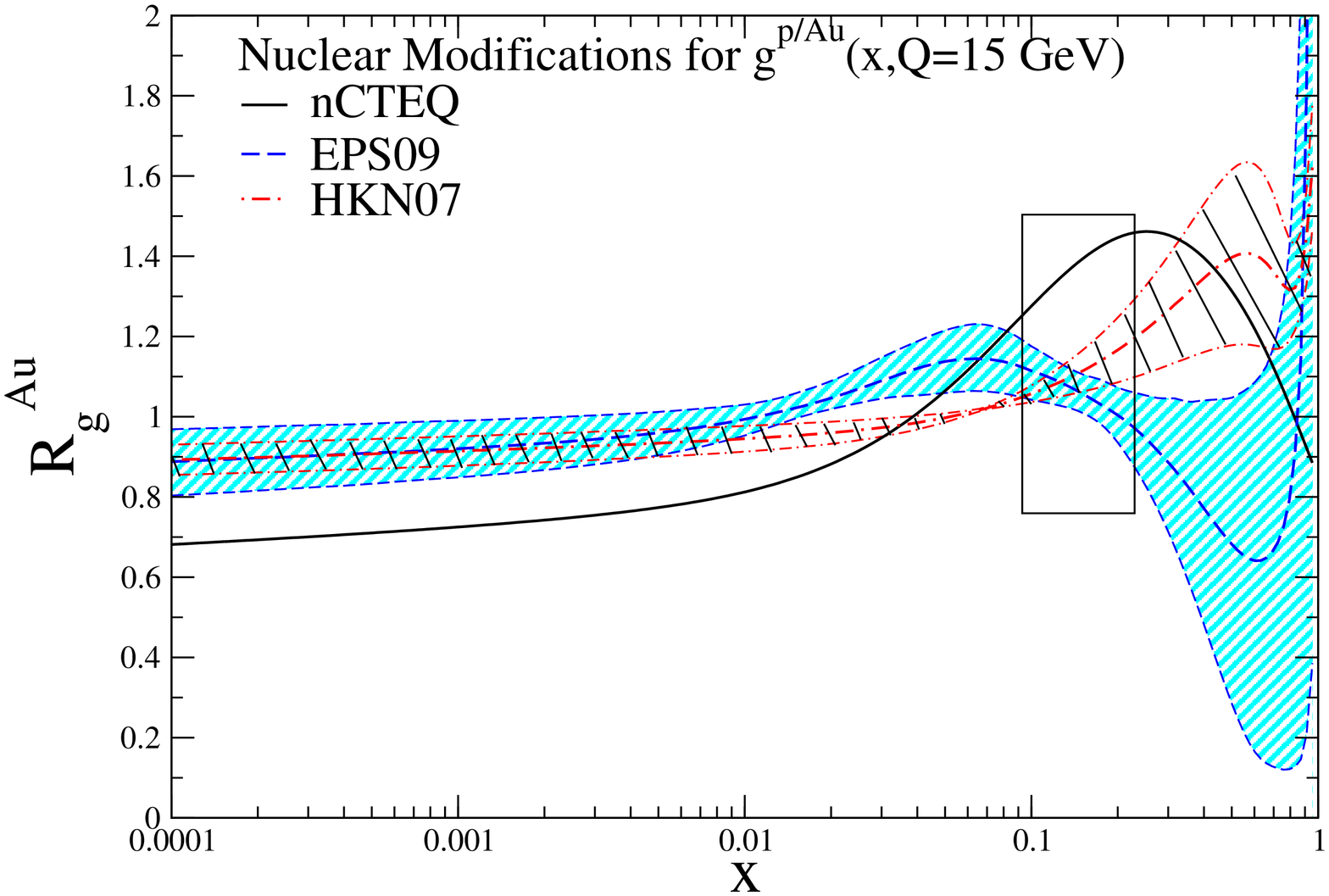}
\includegraphics[angle=-90,scale=0.26]{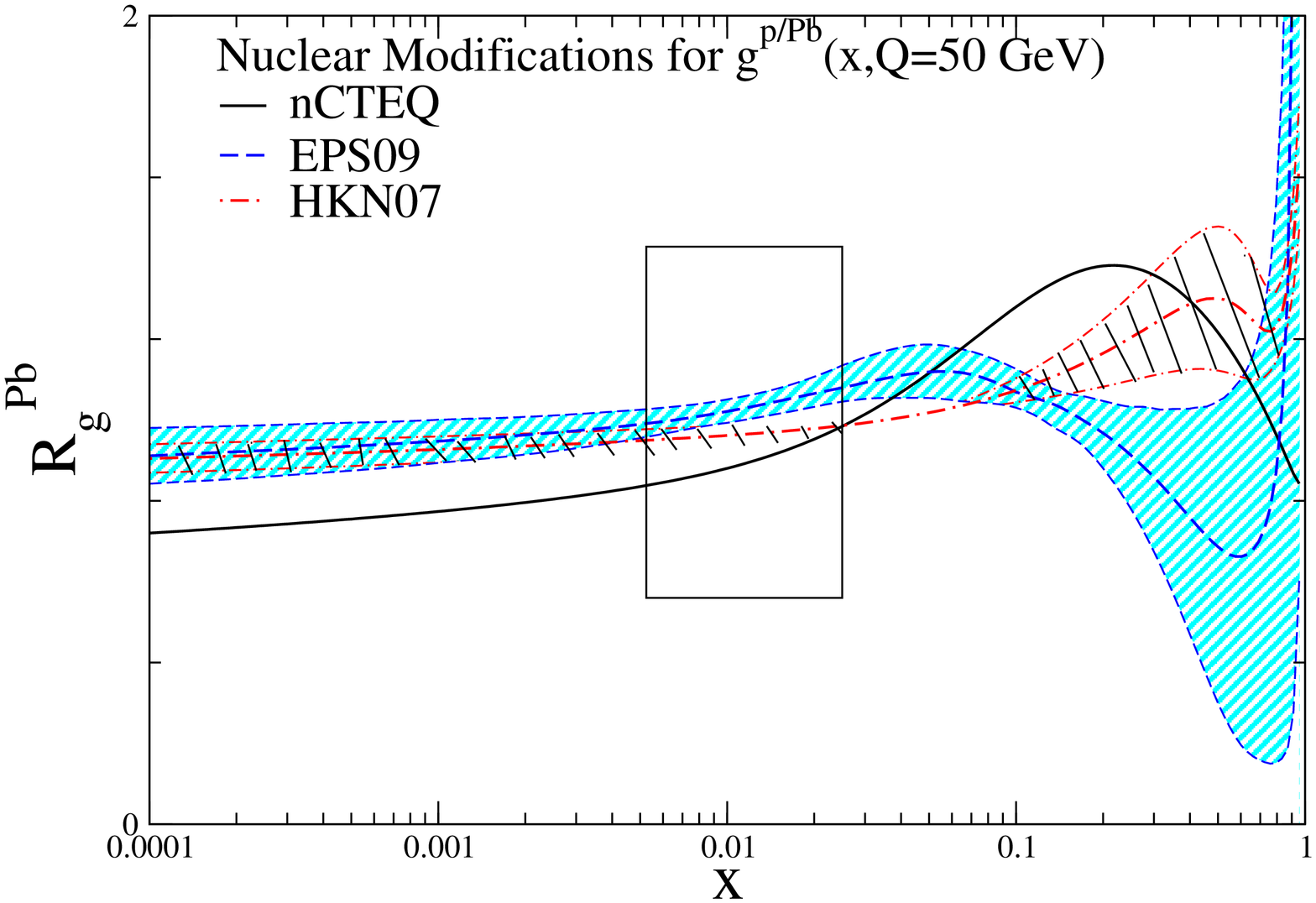}
\caption{Nuclear modifications ${R_g^{A}}=g^{p/A}(x,Q)/g^{p}(x,Q)$.
Left: for gold at $Q=15$ GeV. Right: for lead at $Q=50$ GeV.
Shown are results for nCTEQ decut3 (solid, black line),
EPS09 (dashed, blue line) + error band, HKN07 (dash-dotted, red line) + error band.
The boxes exemplify the $x$-regions probed at RHIC ($\sqrtsnn=200$~GeV)
and the LHC ($\sqrtsnn=8.8$~TeV), respectively.}
\label{fig:gluon}
\end{center}
\end{figure}

In order to compare the various nPDF sets, we plot in Fig.~\ref{fig:gluon} 	
the gluon distribution ratio $R_g^{A}(x, Q)=g^{p/A}(x, Q)/g^p(x, Q)$ 
as a function of $x$ for a gold nucleus at $Q=15$ GeV (left)
and for a lead nucleus at $Q=50$ GeV (right).
The chosen hard scales $Q=15, 50$ GeV are typical for prompt photon production at RHIC and
the LHC, respectively, and the boxes highlight the $x$-regions probed by these colliders.

As can be seen, the nuclear gluon distribution is very poorly constrained\footnote{Note 
also that at lower scales the uncertainties of the nPDFs are even more pronounced.},
especially in the regions $x<0.02$ and $x>0.1$.
The uncertainty bands of the HKN07 and EPS09 gluon distributions do not overlap 
for a wide range of momentum fractions with $x>0.02$.
Also the rather narrow and overlapping bands at small $x<0.02$ do not reflect any constraints
by data, but instead are theoretical assumptions imposed on the small-$x$ behavior of the gluon distributions.
The nCTEQ gluon has again quite a different $x$-shape which is considerably larger (smaller) in 
the $x$-region probed by RHIC (the LHC) as compared to HKN07 and EPS09.

\begin{table}[t]
\begin{center}
 \begin{tabular}{|c|c|c|c|}
\hline 
\textbf{\scriptsize Name}  & (initial) fit parameter  & $c_{1,1}$ & $c_{1,2}$ 
\\
\hline
{\tt decut3} & free & -0.29 & -0.09
\\
{\tt decut3g1} & fixed & 0.2 & 50.0
\\
{\tt decut3g2} & fixed & -0.1 & -0.15
\\
{\tt decut3g3} & fixed & 0.2 & -0.15
\\
{\tt decut3g4} & free & 0.2 & -0.15
\\
{\tt decut3g5} & fixed & 0.2 & -0.25
\\
{\tt decut3g7} & fixed & 0.2 & -0.23
\\
{\tt decut3g8} & fixed & 0.35 & -0.15
\\
{\tt decut3g9} & fixed -- free proton & 0.0 & ---
\\
\hline
\end{tabular}
\end{center}
\caption{\label{tab:decutg}Start values for the parameter $c_1 = c_{1,0} + c_{1,1} (1- A^{-c_{1,2}})$ 
governing the small $x$ behavior of the gluon distribution 
at the initial scale $Q_0=1.3$~GeV. The parameter $c_{1,0}$ corresponds to the gluon in the proton and
has been kept fixed. 
With one exception, {\tt decut3g4}, the parameters $c_{1,1}$ and $c_{1,2}$ have been kept fixed as well.
For further details on the functional form the reader may refer to Ref.~\protect\cite{Schienbein:2009kk}.
}
\end{table}

At present, the nCTEQ nPDFs do not come with an error band.
In order to assess the uncertainty of the nuclear gluon PDF 
we have performed a series of global fits to $\ell A$ DIS and Drell-Yan data
in the same framework as described in Ref.~\cite{Schienbein:2009kk}.
However, each time we have varied assumptions on the functional form of the 
gluon distribution\footnote{The corresponding sets of nPDFs are available upon request from the authors.}.
More precisely, the coefficient $c_1 = c_{1,0} + c_{1,1} (1- A^{-c_{1,2}})$ influencing the
small $x$ behavior of the gluon distribution, see Eq.~(1) in \cite{Schienbein:2009kk}, has been varied
as summarized in Table~\ref{tab:decutg}. 
Each of these fits is equally acceptable with an excellent $\chi^2$/dof in the range of $\chi^2$/dof$=0.88$--$0.9$.

\begin{figure*}
\begin{picture}(480,140)(0,0)
\put(0,145){\mbox{\includegraphics[angle=-90,scale=0.26]{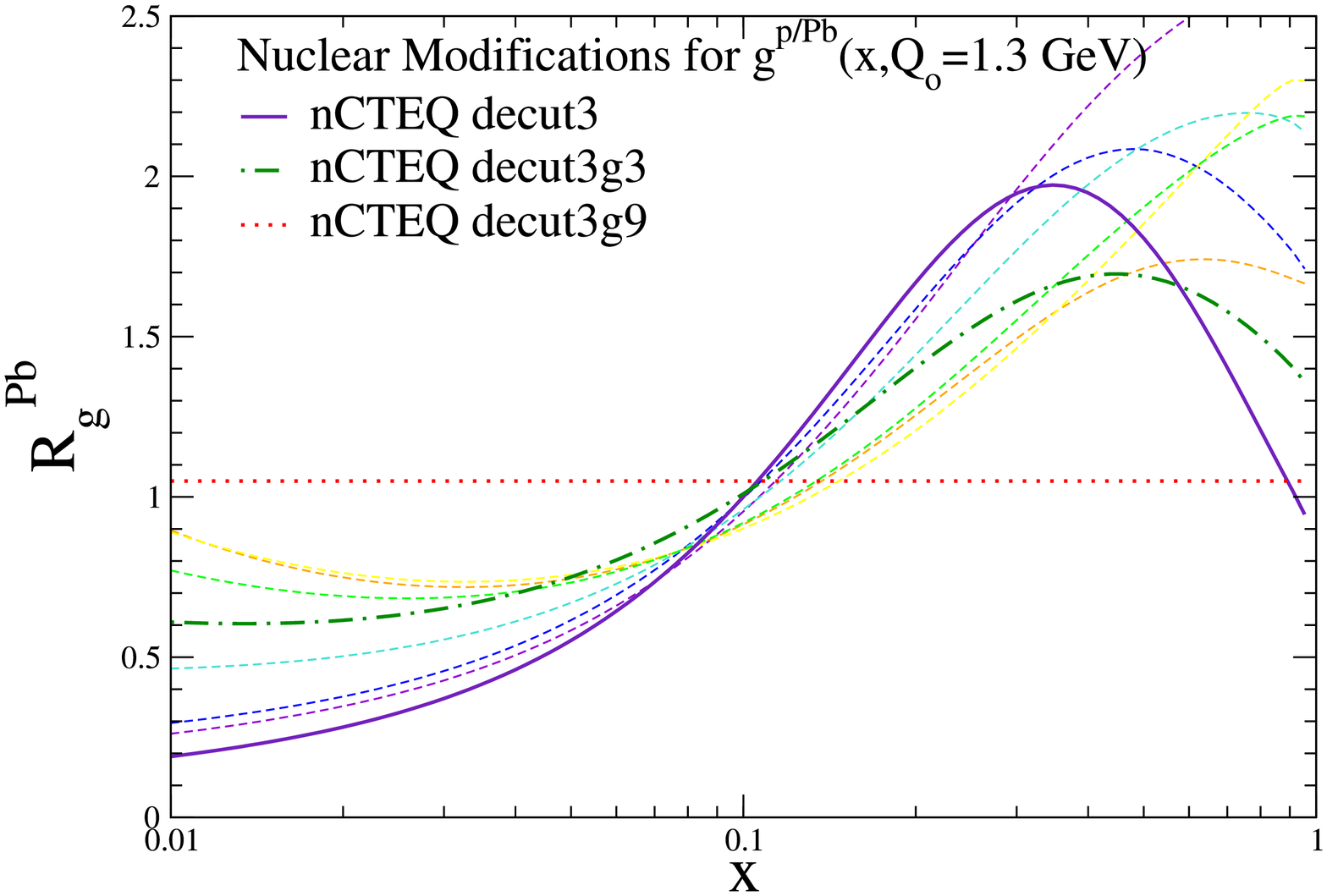}}}
\put(220,2){\mbox{\includegraphics[angle=0,scale=0.40]{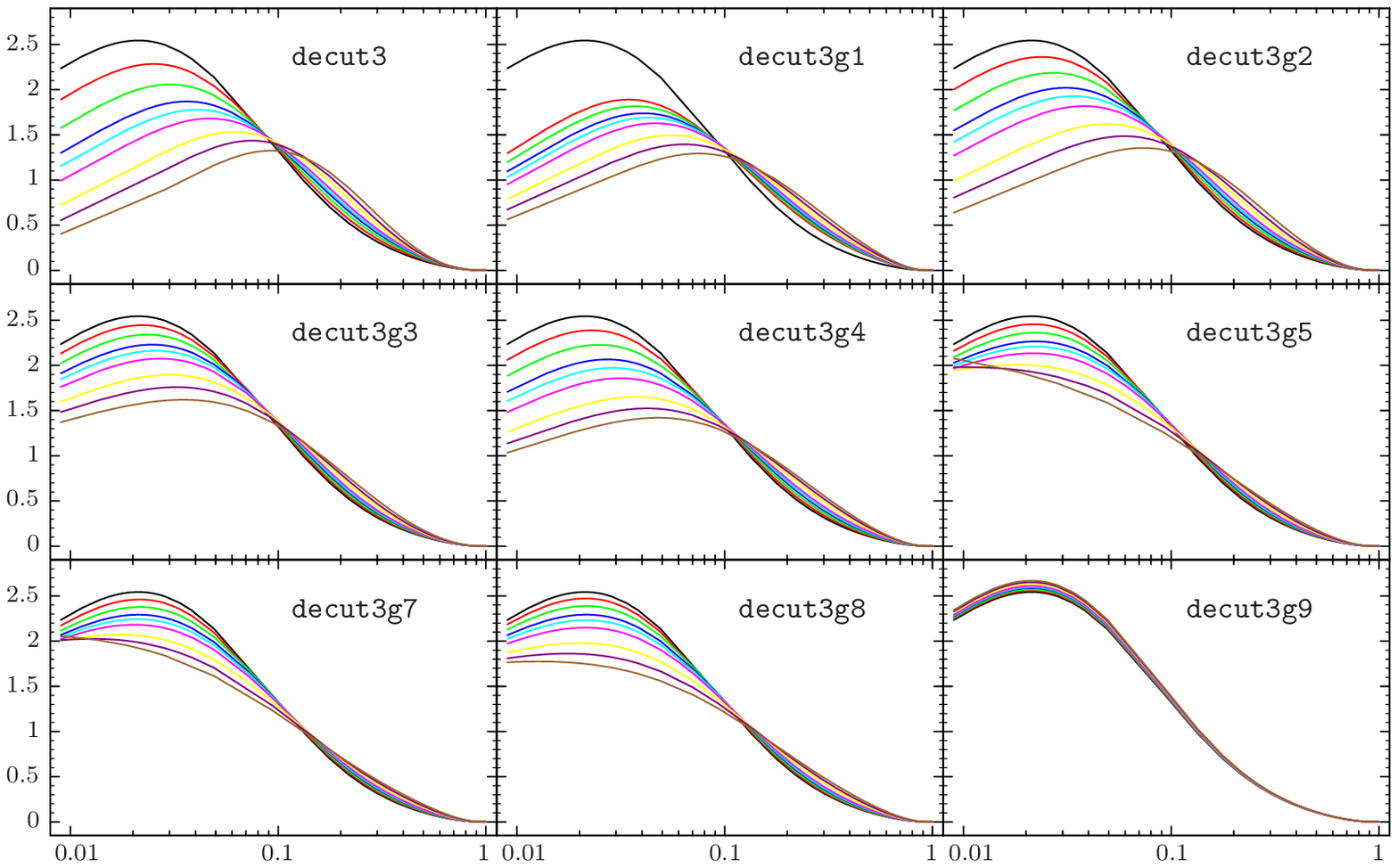}}}
\put(310,-2){\small{x}} 
\put(210,50){\rotatebox{90}{\small{$xg^{p/A}(x,Q_0)$}}} 
\end{picture}
\caption{Left: nPDF ratio $R_g^{Pb}$ at $Q_0=1.3$~GeV predicted within the different nCTEQ sets -- fits from top to bottom: {\tt decut3g9, decut3g5, decut3g7, decut3g8, decut3g3, decut3g4, decut3g2, decut3g1, decut3g}. Right: nCTEQ gluon nPDFs for different A (1, 2, 4, 9, 12, 27, 56, 108, 207) vs x at $Q_0=1.3$~GeV -- from left to right, and top to bottom: {\tt decut3g, decut3g1, decut3g2, decut3g3, decut3g4, decut3g5, decut3g7, decut3g8, decut3g9}.}
\label{fig:decut3gx2}
\end{figure*}

In order to give an idea about the gluon nPDF uncertainty, we plot in Fig.~\ref{fig:decut3gx2} (left)
a collection of ratios $R_g^{Pb}$ for a lead nucleus as 
a function of the momentum fraction $x$ at the initial scale $Q_0=1.3$~GeV, while in Fig.~\ref{fig:decut3gx2} (right) the actual 
gluon nPDFs are plotted versus $x$ for a range of $A$ values.  
Results are shown for several of the fits of the \texttt{decut3g} series.
The ensemble of these curves together with the HKN07 and EPS09 uncertainty bands provides a much more
realistic estimate of the uncertainty of the nuclear gluon distribution which is clearly
underestimated by just one individual error band.
This is due to the fact that for a specific fit, assumptions on the functional form of the nPDFs have been
made so that the error bands based on the Hessian matrix for a given minimum only reflect the uncertainty 
relative to this set of assumptions.

In order to explore the allowed range of nCTEQ predictions for the nuclear production ratios to be 
discussed in Sec.~\ref{sec:pA_RHIC} and \ref{sec:pA_LHC}  we choose the three sets
\texttt{decut3} (solid black line), \texttt{decut3g9} (dotted red line), and \texttt{decut3g3} 
(dash-dotted green line).  The original fit \texttt{decut3}~\cite{Schienbein:2009kk} exhibits a very strong shadowing at small $x$;
conversely, the \texttt{decut3g9} fit closely follows the distribution of the gluon in a (free) proton
and the \texttt{decut3g3} gluon lies between the two extremes.
In most cases, however, we focus on the original fit \texttt{decut3}
to which we refer by default as nCTEQ, if the fit name is not specified.
Together, with the HKN07 and EPS09 predictions this will cover to a good degree the range of possibilities for
the nuclear production ratios.

\begin{figure} 
\begin{center}
\includegraphics[angle=-90,scale=0.26]{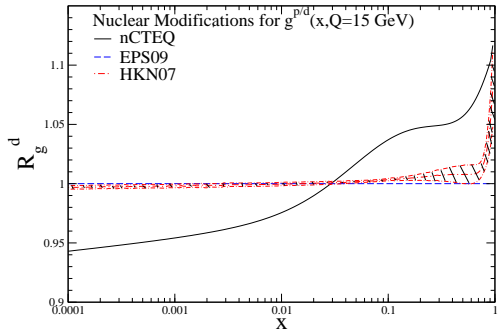}
\caption{Nuclear modifications to deuteron, ${R_g^{d}}=g^{p/d}(x,Q)/g^{p}(x,Q)$ at $Q=15$ GeV, 
nCTEQ (solid black line), EPS09 (dashed blue line), HKN (dash-dotted red line) + error band}
\label{fig:nPDFRHIC}
\end{center}
\end{figure}

At RHIC, the incoming projectile is not a proton but a deuteron nucleus ($A=2$), whose PDFs may be 
different from that of a proton. In Fig.~\ref{fig:nPDFRHIC} the expected nuclear modifications of 
the deuteron nucleus are shown.  
The EPS09 nPDFs do not include nuclear corrections to the deuteron PDFs, while the HKN and nCTEQ sets do.  
Those corrections are not large, at most 5\%, with nCTEQ having them more pronounced.  

\subsection{Heavy-quark sector}

\begin{figure}[t]
\begin{center}
\includegraphics[scale=0.26,angle=-90]{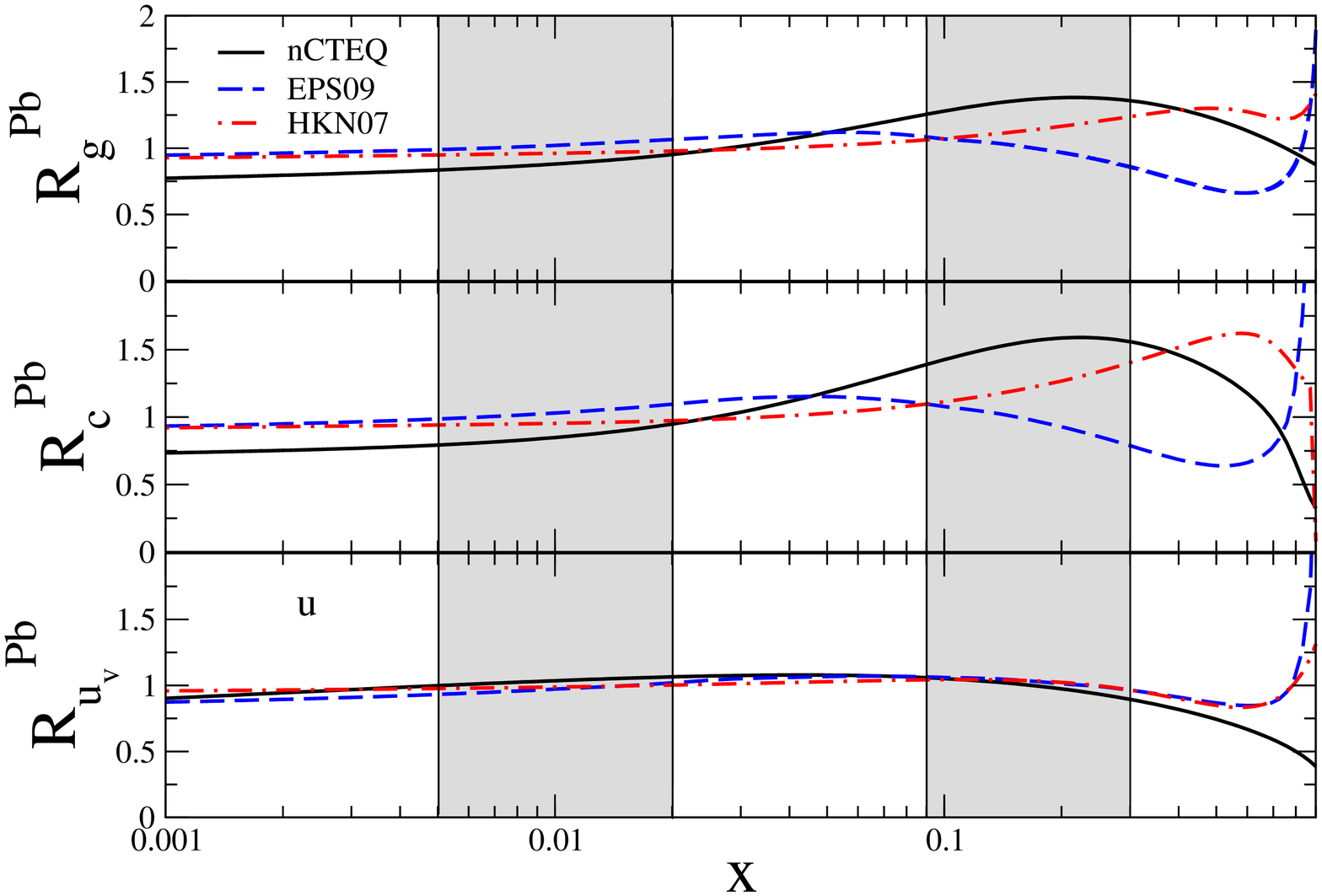}
\includegraphics[scale=0.26,angle=-90]{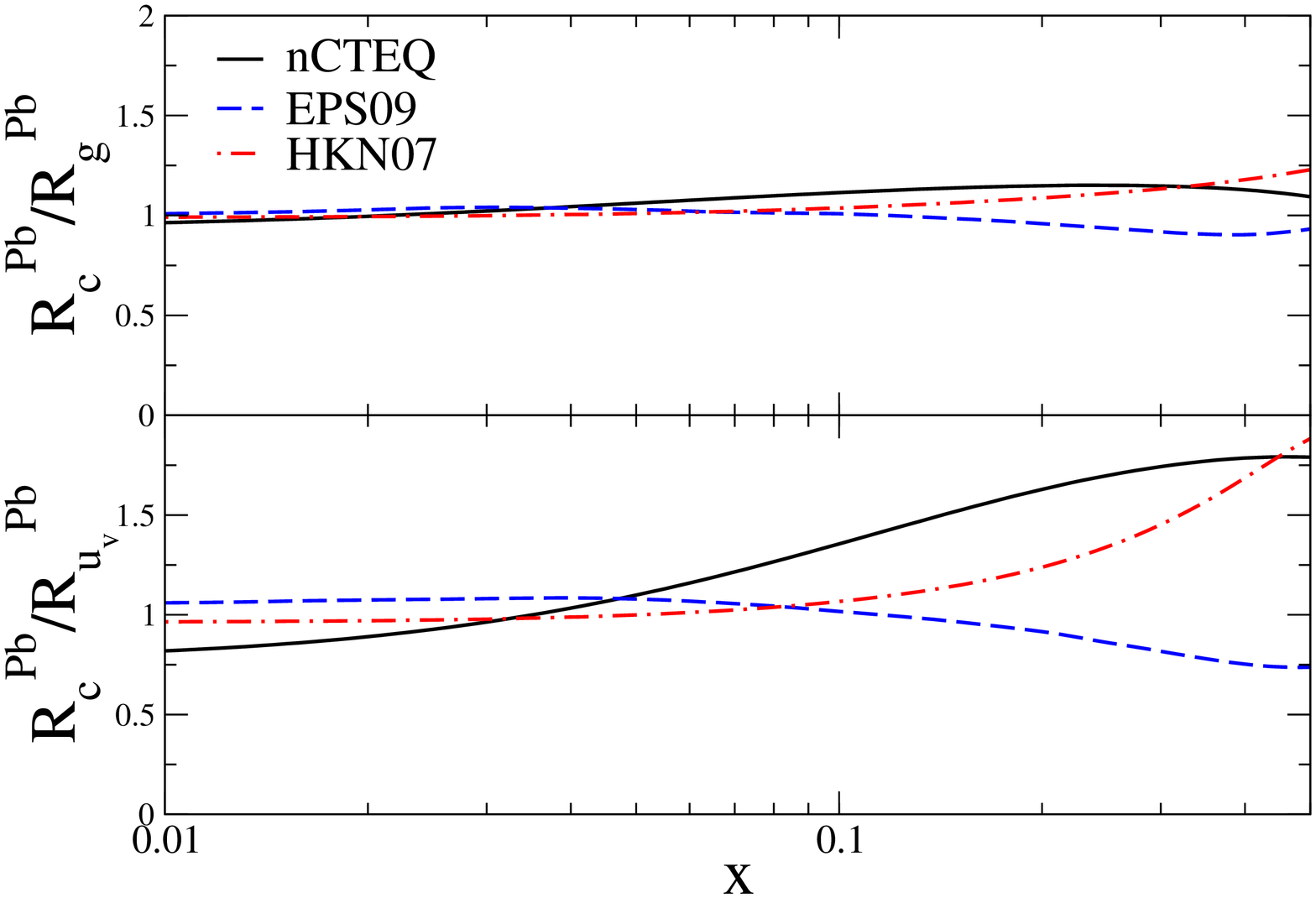}
\caption{Left: nPDF ratios $R_g^{Pb}=g^{p/Pb}(x,Q)/g^{p}(x,Q)$ (top) , 
$R_c^{Pb}=c^{p/Pb}(x,Q)/c^{p}(x,Q)$ (middle), 
$R_{u_v}^{Pb}=u_v^{p/Pb}(x,Q)/u_v^{p}(x,Q)$ (bottom) 
at $Q=50$ GeV within nCTEQ (solid black line), 
EPS09 (dashed blue line), and HKN07 (dash-dotted red line). 
The shaded regions correspond to the $x$-values probed at RHIC ($x\sim 10^{-1}$) and the 
LHC ($x\sim 10^{-2}$). 
Right: double ratios $R_c^{Pb}/R_g^{Pb}$ and $R_c^{Pb}/R_{u_v}^{Pb}$ 
using the same nPDF sets.}
\label{fig:ratio_gcu}
\end{center}
\end{figure}

Let us now turn to the heavy-quark distribution. 
In the standard approach used in almost all global analyses of PDFs, the heavy-quark distributions are generated
radiatively, according to DGLAP evolution equations~\cite{Altarelli:1977zs,Gribov:1972ri,Dokshitzer:1977sg}, 
starting with a perturbatively calculable boundary condition 
\cite{Collins:1986mp,Buza:1998wv} at a scale of the order of the heavy-quark mass.
In other words, there are no free fit parameters associated to the heavy-quark distribution and
it is entirely related to the gluon distribution function at the scale of the boundary condition. 
As a consequence, the nuclear modifications to the radiatively generated heavy-quark PDF are very similar to
those of the gluon distribution\footnote{Note that in Mellin moment $N$-space the relation $c^{p/A}(N,Q)/c^p(N,Q) \simeq 
g^{p/A}(N,Q)/g^p(N,Q)$ holds approximately for $Q \sim Q_0$.} and quite different from the nuclear corrections
in the valence-quark sector.
This feature is illustrated in \mbox{Fig.~\ref{fig:ratio_gcu}} (left) where we show the nuclear modifications for
the gluon (upper panel), charm (middle panel) and the valence up-quark (bottom panel), in a lead nucleus, for three different sets
of nuclear PDFs at the scale $Q=50$ GeV as in Fig.~\ref{fig:gluon} (right).
The shaded regions in Fig.~\ref{fig:ratio_gcu} (left) correspond to the typical $x$-values probed at RHIC ($x\sim 10^{-1}$) and the LHC ($x\sim 10^{-2}$).  
The close similarity between the charm and the gluon nPDFs can be better seen in Fig.~\ref{fig:ratio_gcu} (right) where the {\it double} ratios, $R_c^{Pb}/R_g^{Pb}$ and 
$R_c^{Pb}/R_{u_v}^{Pb}$ ($u_v\equiv u-\bar{u}$ being the valence distribution), are plotted. Remarkably, the nuclear effects in the gluon and the charm PDFs are different 
by at most 20\% at large $x$ ($R_c/R_g\lesssim 1.2$), whereas the difference can be as large as 80\% ($R_c/R_{u_v}\simeq 1.8$) when comparing the valence up-quark and the charm nPDF ratios.
Therefore, in the standard approach, 
the LO direct contribution ($gQ \rightarrow \gamma Q$) only depends on the gluon distribution,
either directly or via the dynamically generated heavy-quark distribution,
making this process an ideal probe of the poorly known gluon nPDF.

Conversely, light-cone models predict a nonperturbative (intrinsic) heavy-quark
component in the proton wave-function~\cite{Brodsky:1980pb,Brodsky:1981se} (see~\cite{Pumplin:2005yf} for an overview of different models).
Recently, there have been studies investigating a possible intrinsic charm (IC) content
in the context of a global analysis of proton PDFs~\cite{Pumplin:2007wg,Nadolsky:2008zw}.  
In the nuclear case, there are no global PDF studies of IC (or IB) available. This is again
mainly due to the lack of nuclear data sensitive to the heavy-quark components in nuclei.
For this reason, we only consider the standard radiative charm approach in the
present paper.
Measurements of $\gamma + Q$ production in $pA$ collisions at backward (forward) rapidities 
are sensitive to the BHPS-IC in nuclei (the proton) complicating the analysis.
A similar statement is true for RHIC,
where due to the lower center-of-mass energy the results depend on the amount of intrinsic charm. 
Therefore, once the nuclear gluon distribution has been better determined from other processes these cases may be useful in the future to constrain the nuclear IC.


\section{Phenomenology at RHIC}
\label{sec:pA_RHIC}
In this section we present the theoretical predictions for the associated production of a photon and 
a heavy-quark jet in $d$--Au collisions at RHIC at $\sqrtsnn=200$~GeV.

\subsection{Cuts}

The experimental cuts used for the theoretical predictions are listed in Table~\ref{table:RHIC-cuts}.  
The photon rapidity and isolation requirements are appropriate for the PHENIX detector\cite{Okada:2005}.  
When $p_{T,\gamma}=p_{T,Q}$ the NLO cross-section is known to become infrared sensitive\footnote{This
back-to-back kinematics matches the LO case and constrains the transverse momentum of the third particle to be zero.}.  
Therefore, in order to acquire an infrared safe cross-section, the minimum transverse momentum of the photon 
is kept slightly 
above that of the heavy-quark~\cite{IRSensitivity:1998,Catani:2002ny} which ensures a
proper cancellation between real and virtual contributions.
Also note that the $p_{T}^{min}$ cuts in the $\gamma + b$ channel ($p_{T,Q}^{min}=14$ GeV and $p_{T,\gamma}^{min}=17$ GeV) were taken to be higher than those in $\gamma + c$ events in order to keep terms of $\cO{m_Q/p_{T}}$ small.

\begin{table}[h]
\begin{center}
\begin{tabular}{c|c|c|c|c}
& $\pt$ & Rapidity & $\phi$& Isolation Cuts  
\\
\hline \hline 
Photon (+c) & $p_{T,\gamma}^{min} = 7$ GeV & $|y_\gamma|<0.35$ & $0^{\circ}<\phi<180^{\circ}$ & $R=0.5$, $\epsilon <0.1E_\gamma$ 
\\
Photon (+b)& $p_{T,\gamma}^{min} = 17$ GeV & $|y_\gamma|<0.35$ & $0^{\circ}<\phi<180^{\circ}$ & $R=0.5$, $\epsilon <0.1E_\gamma$ 
\\
Charm Jet & $p_{T,Q}^{min} = 5$ GeV & $|y_Q|<0.8$ & ------  & ------ 
\\
Bottom Jet & $p_{T,Q}^{min} = 14$ GeV & $|y_Q|<0.8$ & ------ & ------ 
\\
\end{tabular}
\caption{\label{table:RHIC-cuts}Experimental cuts used for the theoretical predictions 
at RHIC.}
\end{center}
\end{table}

\subsection{Spectra and expected rates}
\label{sec:pA_RHIC_rates}

\begin{figure}
\begin{center}
\includegraphics[angle=-90,scale=0.26]{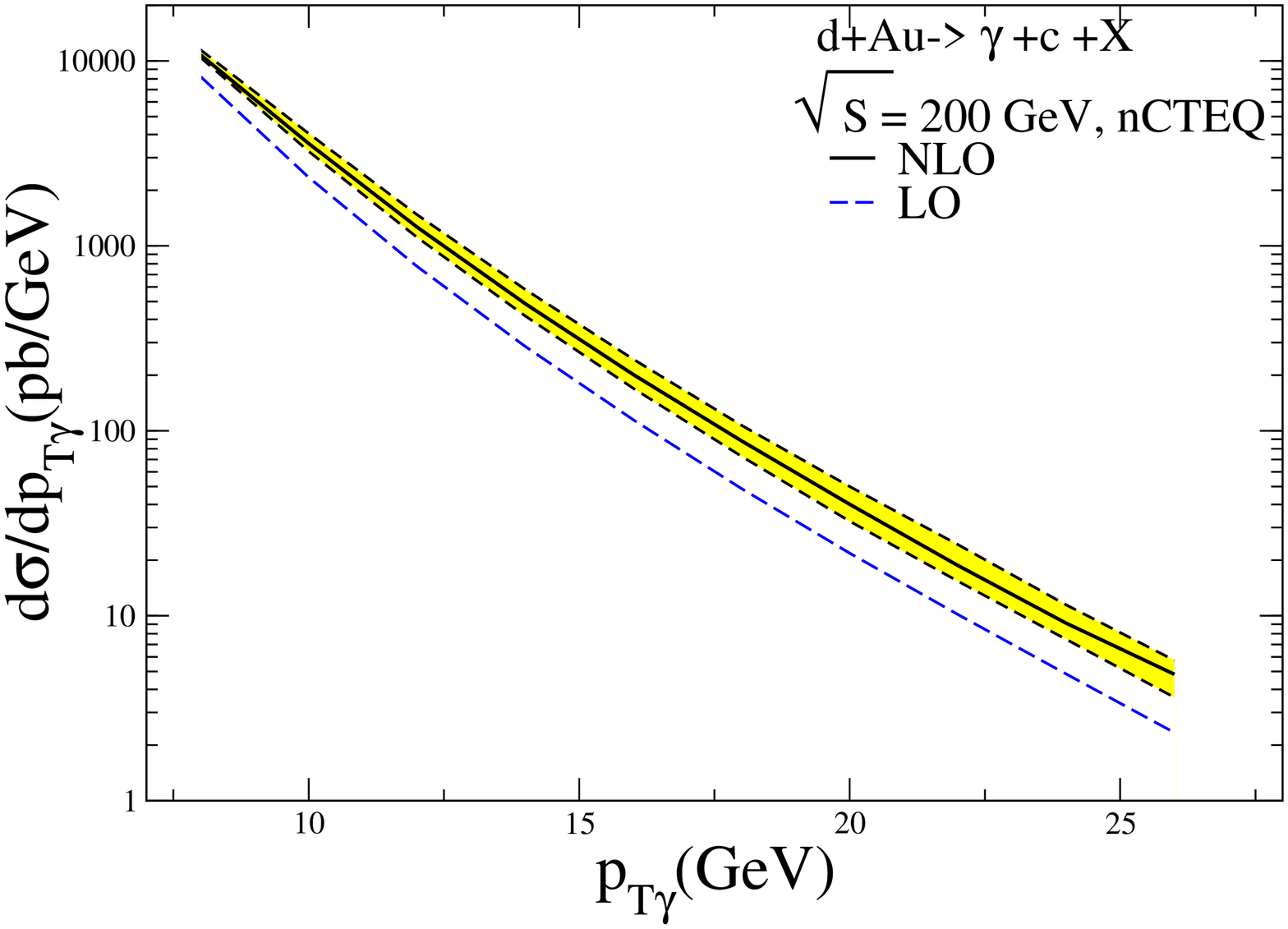}
\includegraphics[angle=-90,scale=0.26]{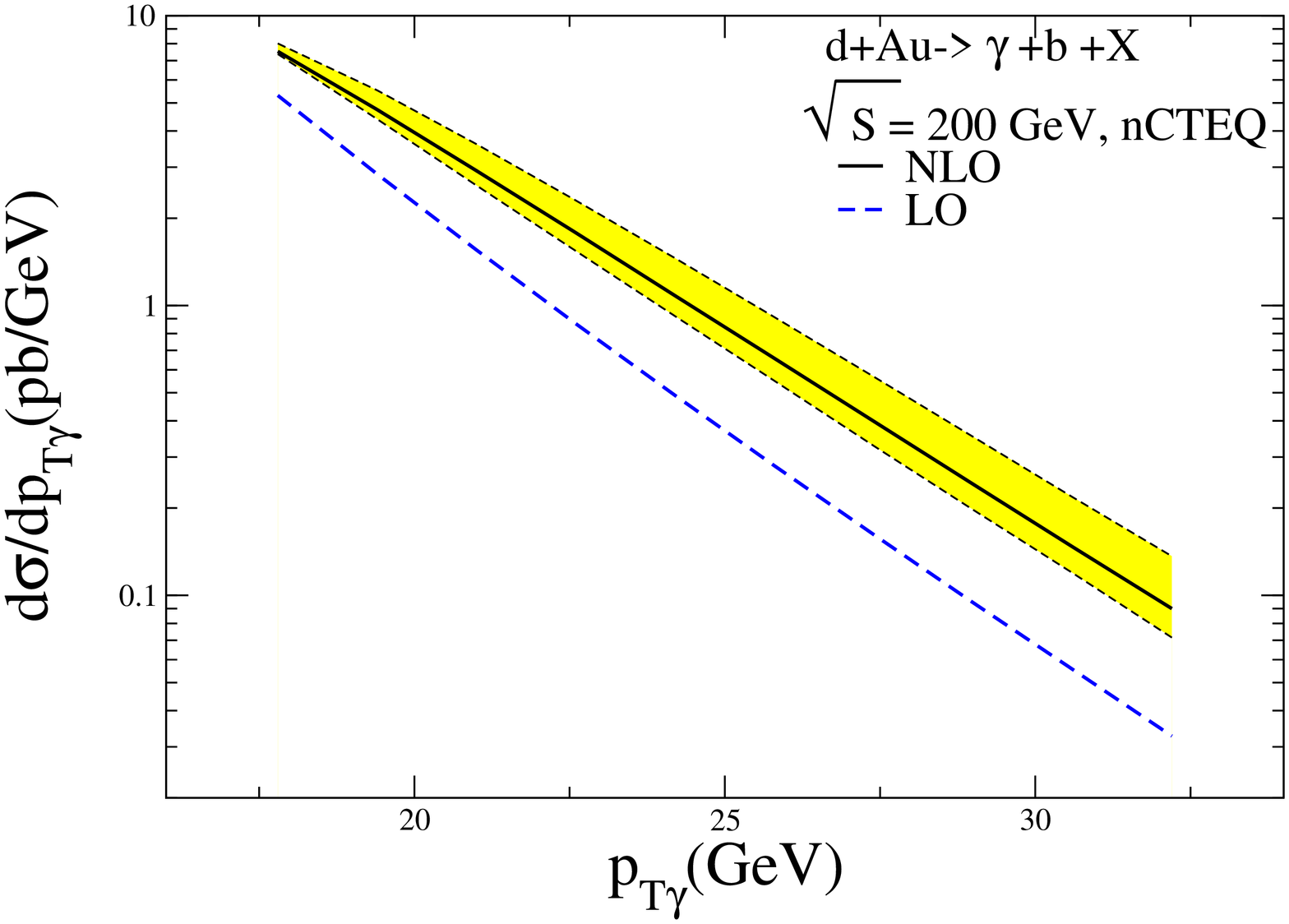}
\caption{Differential cross-section for $\gamma+c$ (left) and $\gamma+b$ (right) production in $d$--Au collisions at a center-of-mass energy of $\sqrtsnn=200$ GeV: NLO (solid black line + band), LO (dashed blue line).}
\label{fig:cross-sec2RHIC}
\end{center}
\end{figure}

The $p_{T_\gamma}$ spectra are shown for $\gamma + c$ production in Fig.~\ref{fig:cross-sec2RHIC} (left) 
and $\gamma + b$ production in Fig.~\ref{fig:cross-sec2RHIC} (right) where the band 
represents the scale uncertainty obtained by varying 
the renormalization, factorization and fragmentation  scales by a factor of two around the
central scale choice, i.e., $\mur=\mui=\mu_f=\xi p_{T\gamma}$ with $\xi = 1/2, 2$.

The total integrated cross-section for $\gamma + c$ events is 
$\sigma^{dAu}_{\gamma+c}=37036$~pb.  
Using the projected weekly luminosity for $d Au$ collisions at RHIC-II,  
${\cal L}^{week}=62$~nb$^{-1}$~\cite{RHIC:WorkReport}, and assuming 12 weeks of ion runs per year, 
the yearly luminosity is ${\cal L}^{year}=744$~nb$^{-1}$.  
Thus, an estimate of the number of events expected in one year is
$N_{\gamma+c}^{dAu}={\cal L}^{year}\times\sigma^{dAu}_{\gamma+c}\simeq 2.8\times10^4$ 
in $d$--Au collisions, without taking into account effects of the experimental acceptances
and efficiencies.
At $p_{T_\gamma}\simeq 20$~GeV (${\rm d}\sigma/{\rm d}p_{T_\gamma}\simeq45$~pb/GeV), the number of events 
would still be large, $\cO{10^2}$ per GeV-bin. 
This indicates that the number of $\gamma$+c events in a year produced at RHIC-II will be substantial. 
The rates expected in the $\gamma+b$ channel at RHIC are naturally much more modest. 
Using the total integrated cross-section $\sigma^{dAu}_{\gamma+b}=32$~pb, the number of events to 
be expected in a year is $N_{\gamma+b}^{dAu}=24$. 
Therefore we shall mostly focus the discussion on the $\gamma+c$ channel in the following.

\begin{figure}[t]
\begin{center}
\includegraphics[angle=-90,scale=0.28]{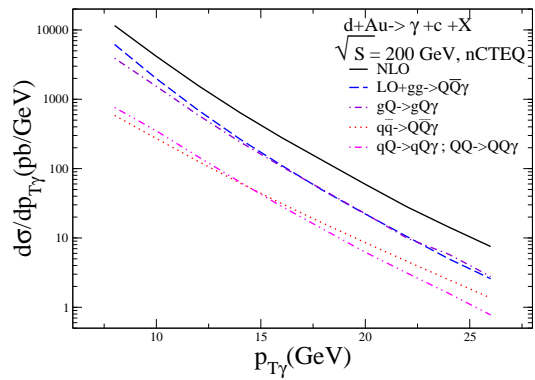}
\caption{Subprocess contributions to the differential cross-section at RHIC: NLO (solid black line), 
LO$+ gg \rightarrow Q\bar Q\gamma$ (dashed blue line), 
$gQ\rightarrow gQ\gamma$ (dash-dotted purple line), $q\bar q \rightarrow Q\bar Q \gamma$ (dotted red line), 
$q(\bar q) Q(\bar Q) \rightarrow q(\bar q) Q(\bar Q) \gamma,QQ \rightarrow QQ \gamma$ (dash-dot-dotted magenta line).}
\label{fig:K-factorRHIC}
\end{center}
\end{figure}

In Fig.~\ref{fig:K-factorRHIC} the individual subprocess contributions to the $\gamma+c$ NLO production cross-section are presented.   
As can be seen, the dominant subprocesses are the LO Compton scattering $gQ\rightarrow \gamma Q$, as well as the higher-order $gQ\rightarrow \gamma gQ$ and $gg\rightarrow \gamma Q \bar Q$ channels.  
Thus almost all the PDF dependence in the NLO $\gamma+c$ cross-section comes from the gluon and 
heavy-quark PDFs and not from the light-quark PDFs.
The relative increase of the contributions by the annihilation subprocess, 
$q \bar q \rightarrow Q \bar Q \gamma$, and the light quark-heavy quark subprocess  $qQ\rightarrow qQ \gamma$ at higher $x$ ($p_{T\gamma} \sim 15$~GeV)
is due to the slower decrease of the 
valence quark PDF at high $x$ as compared to the rest of the PDFs.  

\begin{figure}
\begin{center}
\includegraphics[angle=-90,scale=0.26]{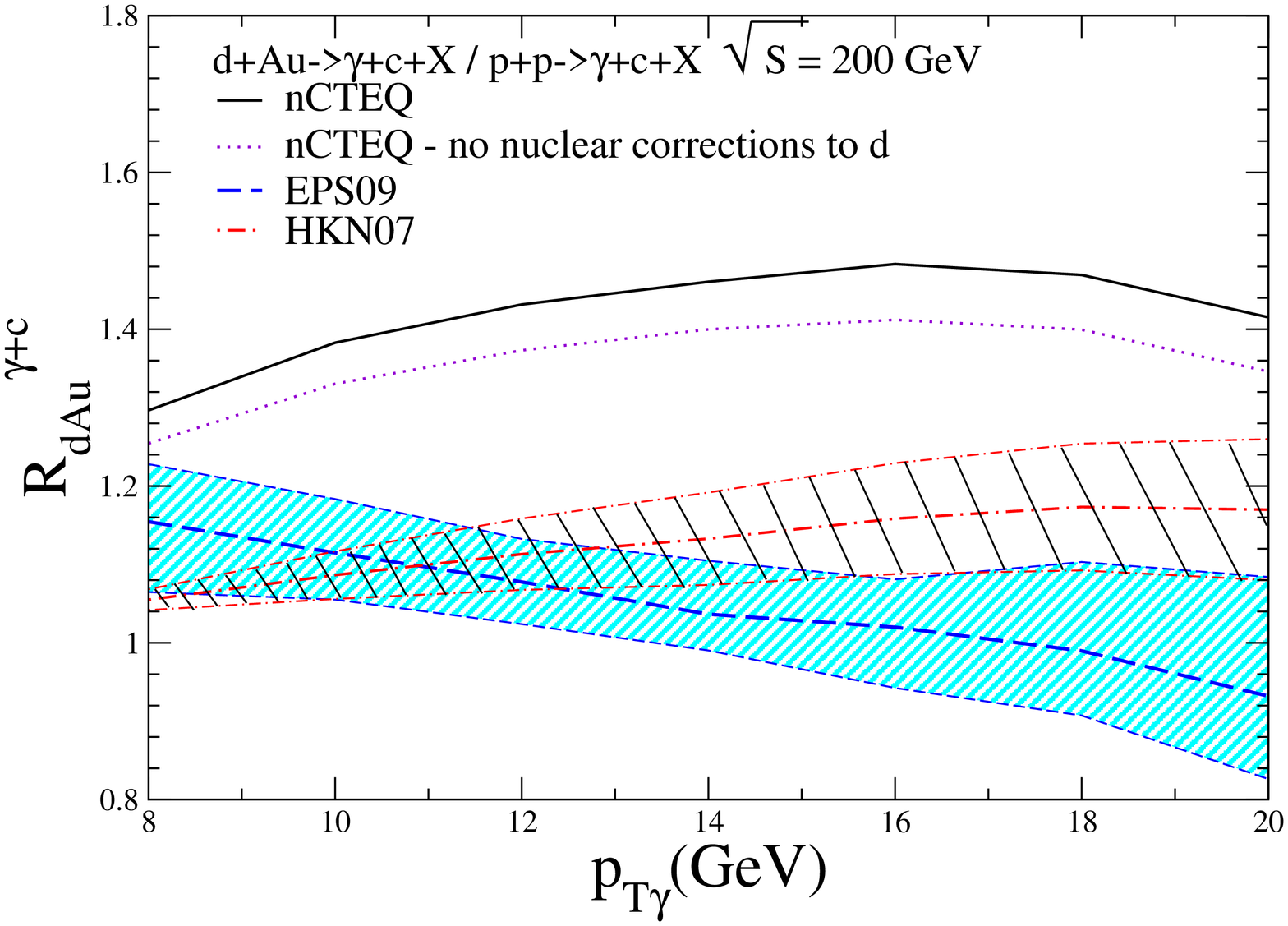}
\includegraphics[angle=-90,scale=0.26]{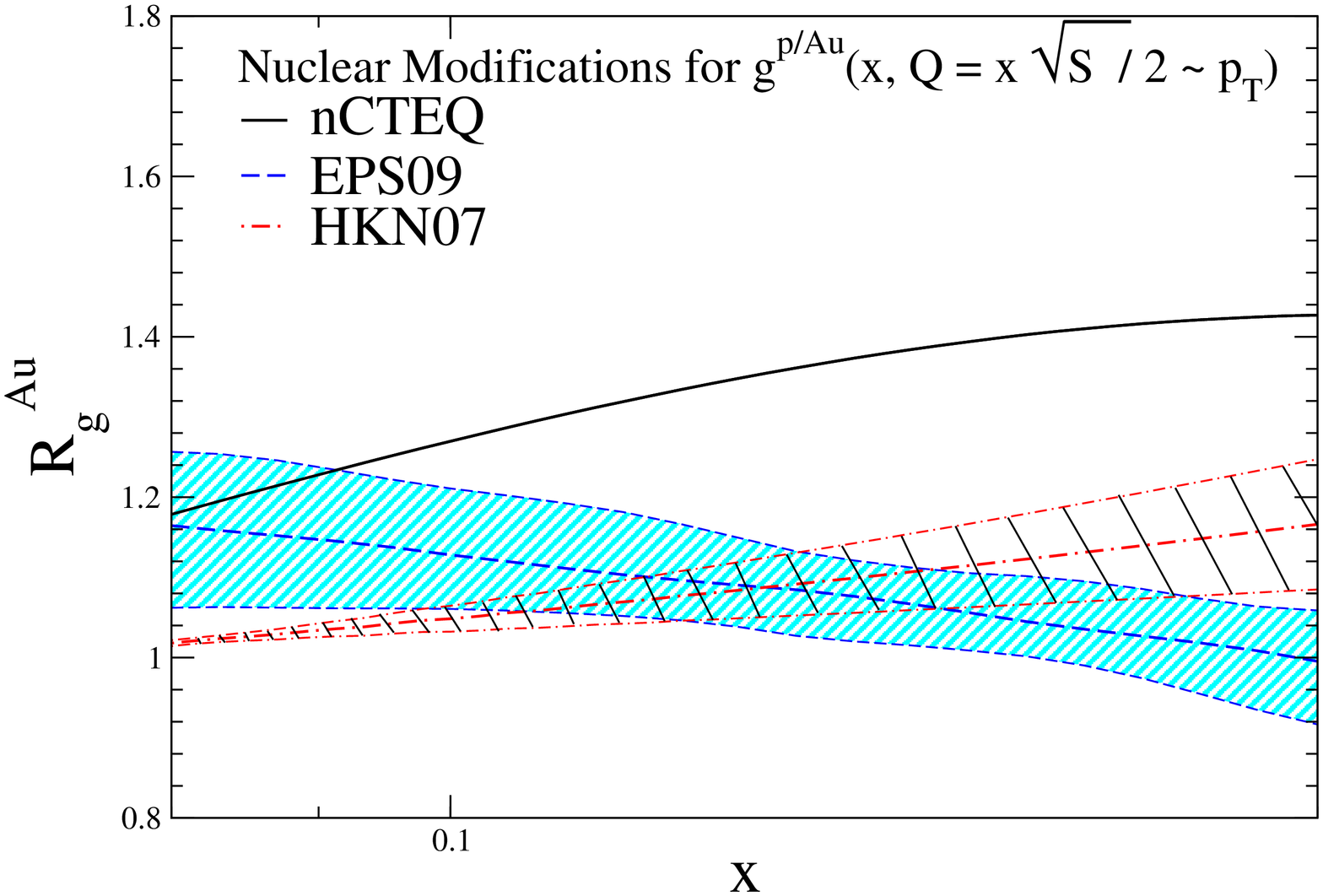}
\caption{Left: nuclear production ratio of the $\gamma+c$ cross-section at RHIC using nCTEQ (solid black line), 
nCTEQ without nuclear corrections in the deuteron (dotted magenta line), EPS09 (dashed blue line) + error band, 
HKN (dash-dotted red line) + error band. 
Right: nuclear modification of the gluon in gold, $R_g^{Au}(x,Q=x\sqrt{S}/2 \sim p_T)$, for the $x$-region probed at RHIC.  This figure corresponds to the enlargement of the box region in the left panel of
Fig.~\protect\ref{fig:gluon}.}
\label{fig:ratio_rhic}
\end{center}
\end{figure}

\begin{figure}
\begin{center}
\includegraphics[angle=-90,scale=0.26]{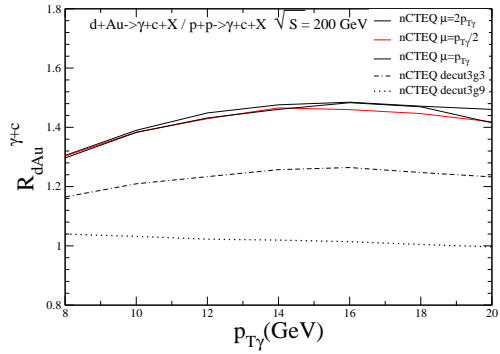}
\caption{
Nuclear production ratio of the $\gamma+c$ cross-section at RHIC 
using the three nCTEQ fits discussed in Sec.~\protect\ref{sec:PDF}. 
Also shown is the scale dependence of $R_{dAu}^{\gamma+c}$.
}
\label{fig:ratio_rhic2}
\end{center}
\end{figure}

\subsection{Nuclear production ratios}

Let us now discuss 
the nuclear modifications of $\gamma + c$ production in $d$--Au collisions. The nuclear production ratio,
\begin{equation}
R^{\gamma+c}_{dAu}
={1\over 2\times197}
{d\sigma/dp_{T\gamma}(d{\rm Au} \rightarrow \gamma+c+X)
\over d\sigma/dp_{T\gamma}(pp \rightarrow \gamma+c+X)},
\end{equation}
is plotted in Fig.~\ref{fig:ratio_rhic} (left) as a function of $p_{T\gamma}$ using the 
three nPDF sets discussed in section \ref{sec:PDF}, namely nCTEQ (solid black line), 
EPS09 (dashed blue line + error band) and HKN (dash-dotted red line + error band).    
There is some overlap between $R^{\gamma+c}_{\rm HKN}$ and 
$R^{\gamma+c}_{\rm EPS09}$ at not too large $\pt\lesssim15$~GeV,
whereas the difference between $R^{\gamma+c}_{\rm nCTEQ}$ on the one hand 
and $R^{\gamma+c}_{\rm HKN}$ 
and $R^{\gamma+c}_{\rm EPS09}$ on the other hand is larger for all transverse momenta. 
The $R^{\gamma+c}_{\rm nCTEQ}$ ratio is further increased by the anti-shadowing corrections 
in the deuteron projectile, 
as can be seen in Fig.~\ref{fig:ratio_rhic} (left) where the nCTEQ predictions are performed with 
(solid line) and without (dashed) corrections in the deuteron (see also Fig.~\ref{fig:nPDFRHIC}).
Due to the rather low center-of-mass energy (as compared to the Tevatron/LHC) the collisions 
at central rapidity at RHIC probe relatively high values of momentum fractions carried by the partons 
in the nuclear target, $x_{_2}=\cO{2\pt/\sqrt{s}}=\cO{10^{-1}}$.  
In Fig.~\ref{fig:ratio_rhic} (right)
we show the nuclear modifications of the gluon distribution in a gold nucleus, 
$R_g^{Au}(x,Q=x\sqrt{S}/2 \sim p_T)$, for the typical $x$-region probed at RHIC. Note that, this figure corresponds to the enlargement of the box-region in the left panel of Fig.~\ref{fig:gluon}.
As can be seen the nuclear production ratios of $\gamma+c$ events
shown in Fig.~\ref{fig:ratio_rhic} (left) 
closely correspond to the different nuclear modifications of the gluon distribution depicted on the 
right side of Fig.~\ref{fig:ratio_rhic}.
Clearly, measurements of this process with appropriately small error bars will be able to distinguish between 
these three different nuclear corrections to the cross-section and therefore be able to constrain the 
gluon nuclear PDF.

In Fig.~\ref{fig:ratio_rhic2} we present the dependence of the nuclear modifications on the 
three nCTEQ fits (\texttt{decut3}, \texttt{decut3g9}, \texttt{decut3g3})
discussed in Sec.~\ref{sec:PDF}.
It is clear that these different fits cover quite a spread of nuclear modifications, 
ranging from ones which are quite pronounced (\texttt{decut3}) to
almost none (\texttt{decut3g9}).
We stress again, that neither of these predictions is preferred over the other
since the nuclear gluon distribution is so poorly known.
Finally, we also show in Fig.~\ref{fig:ratio_rhic2} the scale uncertainty which is entirely 
negligible compared to the PDF uncertainty.

In the next section we present the phenomenology of $\gamma+Q$ production at the LHC
where  smaller values of $x_{_2}$ are probed due to the higher center-of-mass energy.

\section{Phenomenology at LHC}
\label{sec:pA_LHC}

In this section, calculations are carried out for $p$--Pb collisions at the LHC nominal energy, $\sqrtsnn=8.8$~TeV, different from the $pp$ collision energy ($\sqrt{s}=14$~TeV).

\subsection{Cuts}

The cuts used in the present calculation are shown in Table~\ref{table:ALICE-cuts} and are appropriate 
for the ALICE detector\footnote{We have verified that similar results and conclusions are obtained when using 
either ATLAS or CMS acceptances at central rapidities.}~\cite{Conesa:2007zz,Conesa:2007nx,faivre,Abeysekara:2010ze,Aamodt:2008zz}.
Note that the rapidity shown in Table~\ref{table:ALICE-cuts} is given in the laboratory frame, 
which in $pA$ collisions is shifted by $\Delta y=-0.47$ with respect to the center-of-mass frame~\cite{ALICE:volI,ALICE:volII}. In ALICE, photons can be identified in the EMCal electromagnetic calorimeter, or in the PHOS spectrometer with a somewhat more limited acceptance.

\vspace*{1ex}
\begin{table}[h]
\begin{center}
\begin{tabular}{c|c|c|c|c}
& $p_T$ & Rapidity  & $\phi$ & Isolation Cuts  
\\
\hline \hline 
Photon (PHOS) & $p_{T,\gamma}^{min} = 20$ GeV & $|y_\gamma|<0.12$& $220^{\circ}<\phi<320^{\circ}$ & $R=0.2$, $p_T^{\rm th}=2$ GeV
\\
Photon (EMCal) & $p_{T,\gamma}^{min} = 20$ GeV & $|y_\gamma|<0.7$& $80^{\circ}<\phi<180^{\circ}$ & $R=0.2$, $p_T^{\rm th}=2$ GeV
\\
Heavy Jet & $p_{T,Q}^{min} = 15$ GeV & $|y_Q|<0.7$ &  ------ & ------ 
\\
\end{tabular}
\caption{\label{table:ALICE-cuts}\small{Experimental cuts for the ALICE detector.}}
\end{center}
\end{table}

\subsection{Spectra and expected rates}

\begin{figure}
\begin{center}
\includegraphics[angle=-90,scale=0.26]{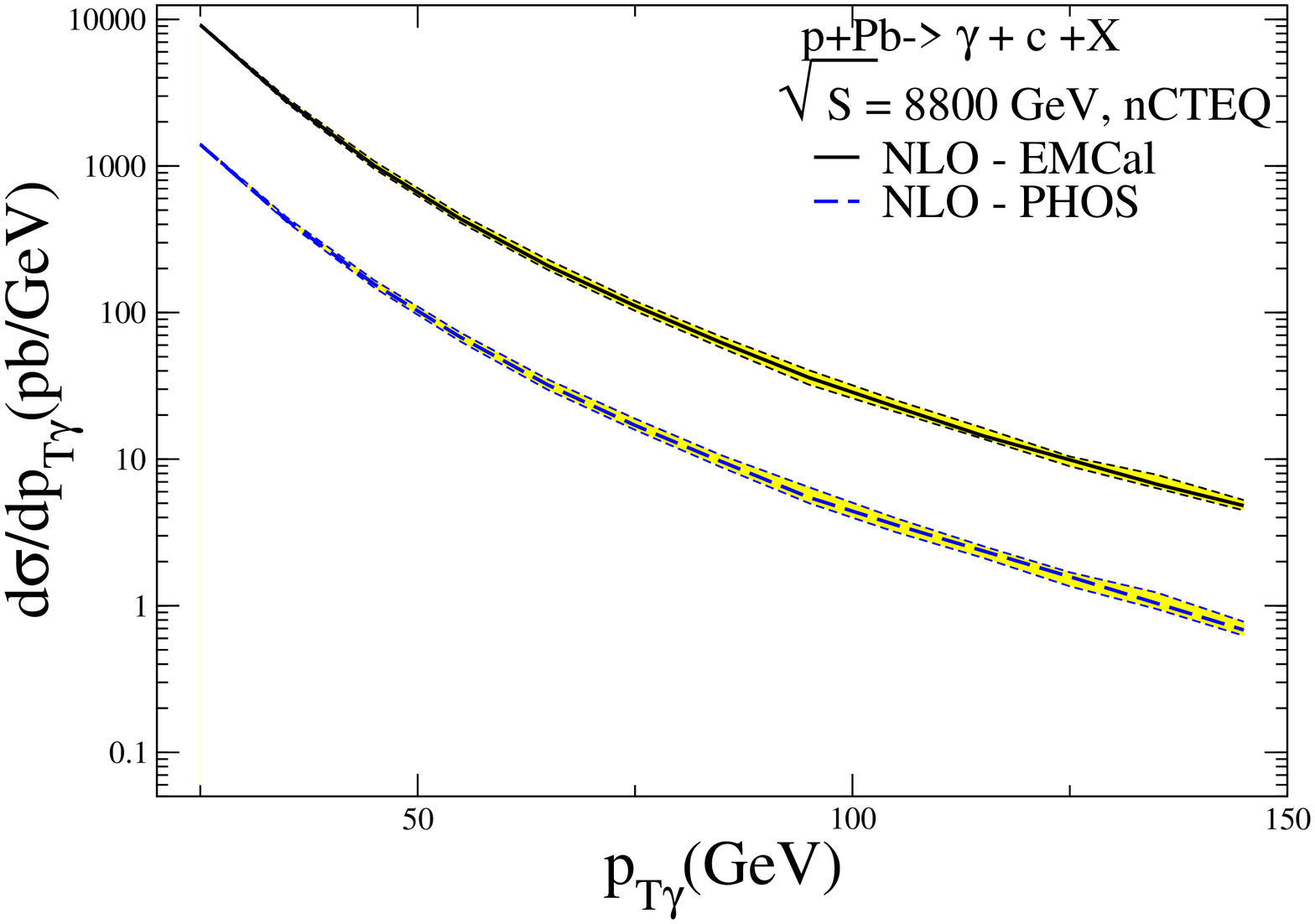}
\includegraphics[angle=-90,scale=0.26]{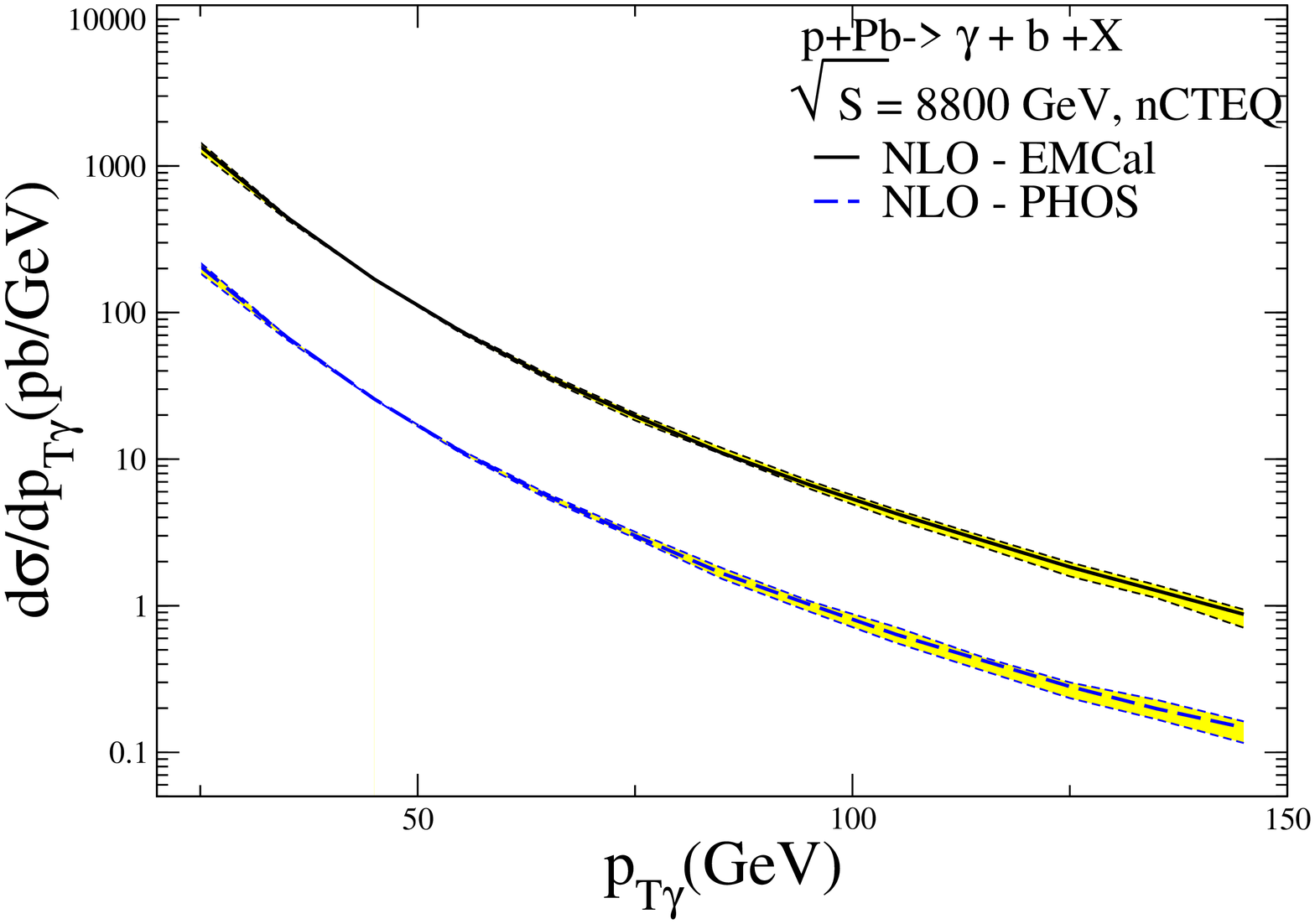}
\caption{NLO differential cross-section for $\gamma+c$ (left) and $\gamma+b$ (right) production in $p$--Pb collisions at a center-of-mass energy of $\sqrtsnn=8.8$ TeV in PHOS (lower band) and EMCal (upper band) acceptances.}
\label{fig:cross-sec2}
\end{center}
\end{figure}

The differential NLO cross-section is plotted as a function of the photon transverse momentum 
in the $\gamma+c$ ($\gamma+b$) channel in 
Fig.~\ref{fig:cross-sec2} left (right) for both PHOS (lower band) and EMCal (upper band); 
the dotted curves indicate the theoretical scale uncertainty.

In order to estimate the number of events produced, we use the instantaneous luminosity
${\cal L}^{inst}=10^{-7}$~pb$^{-1}s^{-1}$~\cite{ALICE:volII} which corresponds to a yearly integrated luminosity of ${\cal L}^{year}=10^{-1}$~pb$^{-1}$ assuming one month ($\Delta t=10^6s$) of running in the heavy-ion mode at the LHC.
In Table~\ref{table:ALICE-sigma} the total integrated cross-section for $\gamma +Q$ for both PHOS and EMCal 
along with the respective anticipated number of events (before experimental efficiencies), $N_{\gamma+Q}^{p Pb}=\sigma^{pPb}_{\gamma+Q}\times {\cal L}^{year}$ are given.  
As expected the $\gamma + b$ and $\gamma + c$ cross-sections at EMCal are increased substantially by the larger acceptance of that detector.  The number of expected $\gamma+b$ events is large, at variance 
with what is expected at RHIC (see section~\ref{sec:pA_RHIC_rates}).

\vspace*{1ex}
\begin{table}[h]
\begin{center}
\begin{tabular}{c|c|c}
& $\sigma^{pPb}_{\gamma+Q}$ & $N^{p Pb}_{\gamma+Q}$ 
\\
\hline \hline 
$\gamma+c$ PHOS & $22700~{\rm pb}$ & 2270
\\
$\gamma+b$ PHOS & $3300~{\rm pb}$ & 330
\\
$\gamma+c$ EMCal & $119000~{\rm pb}$ & 11900
\\
$\gamma+b$ EMCal & $22700~{\rm pb}$ & 2270
\\
\end{tabular}
\caption{\label{table:ALICE-sigma}\small{Total integrated cross-section and number of events 
per year for $\gamma +Q$ production in $p$--Pb collisions at the LHC for PHOS and EMCal acceptances.}}
\end{center}
\end{table}

\begin{figure}
\begin{center}
\includegraphics[angle=-90,scale=0.26]{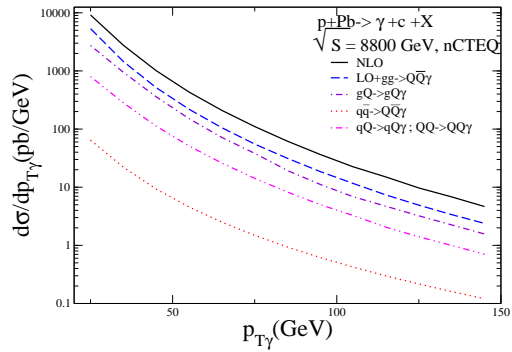}
\caption{Subprocess contributions to the differential cross-section
shown in Fig.~\protect\ref{fig:cross-sec2} (left), NLO (solid black line), LO$+ gg \rightarrow Q\bar Q\gamma$ (dashed blue line), $gQ\rightarrow gQ\gamma$ (dash-dotted purple line), $q\bar q \rightarrow Q\bar Q \gamma$ (dotted red line), 
$qQ \rightarrow qQ \gamma$; $QQ \rightarrow QQ \gamma$ (dash-dot-dotted magenta line).}
\label{fig:cross-sec2_parts}
\end{center}
\end{figure}

The individual subprocess contributions to the cross-section are depicted in Fig.~\ref{fig:cross-sec2_parts}.  
As one can see the Compton ($gQ\rightarrow \gamma Q$) as well as the 
$gQ\rightarrow \gamma gQ$ and $gg\rightarrow \gamma Q \bar Q$ are the dominant subprocesses, demonstrating the sensitivity of this process to the gluon and charm nPDFs.  Here the contribution by the annihilation subprocess proves much smaller than at RHIC.  This is caused by the less pronounced difference in the light anti-quark and heavy quark PDFs at small $x$ as compared to large $x$.  So that now $q \bar q \rightarrow \gamma Q \bar Q$  can no longer compete with the light quark-heavy quark (antiquark) piece of the cross-section ($qQ\rightarrow qQ \gamma$,   
$q\bar Q\rightarrow q \bar Q \gamma$, $\bar qQ\rightarrow \bar qQ \gamma $, $\bar q \bar Q\rightarrow \bar q \bar Q \gamma$).

\begin{figure}
\begin{center}
\includegraphics[angle=-90,scale=0.26]{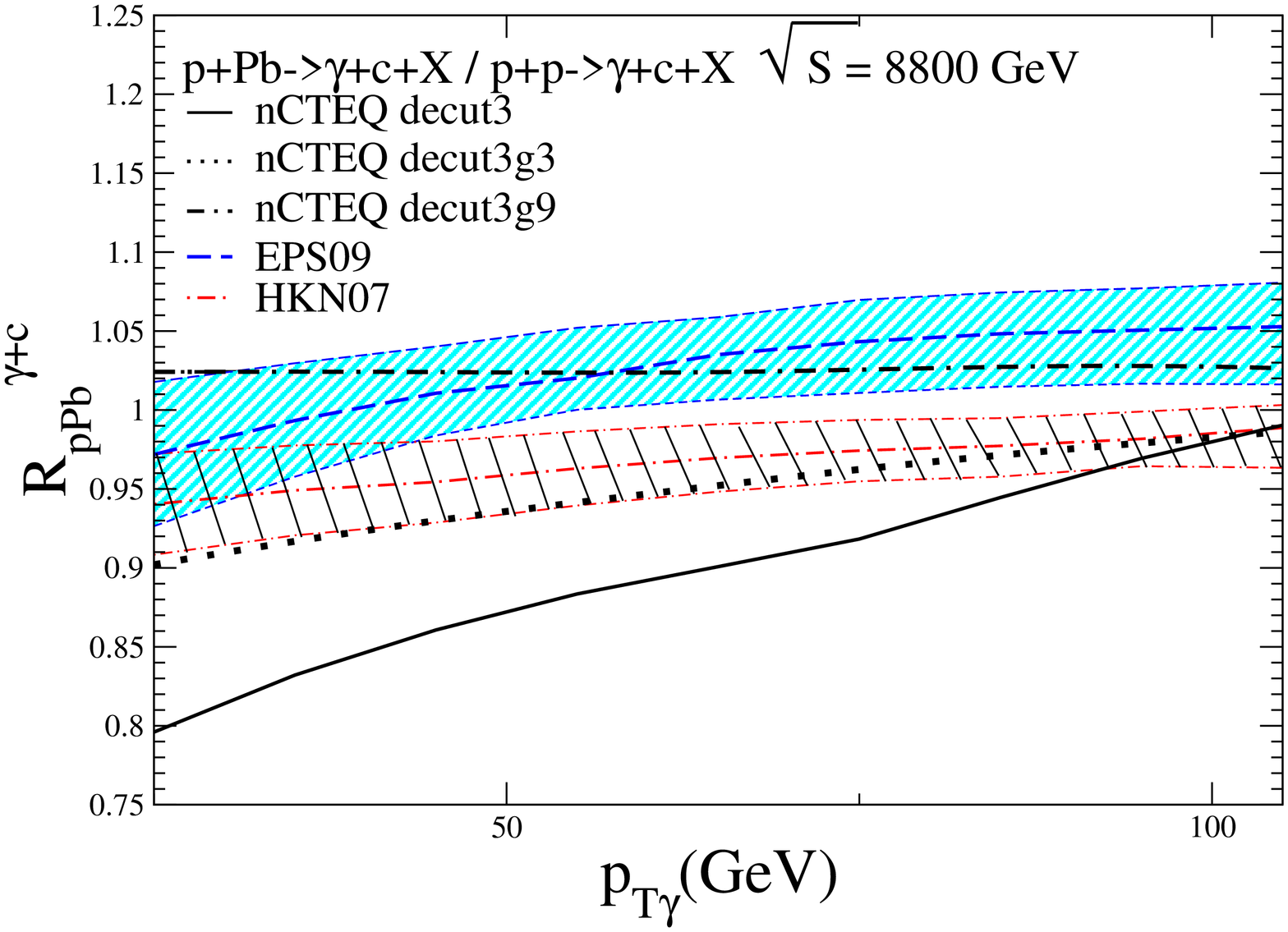}
\includegraphics[angle=-90,scale=0.26]{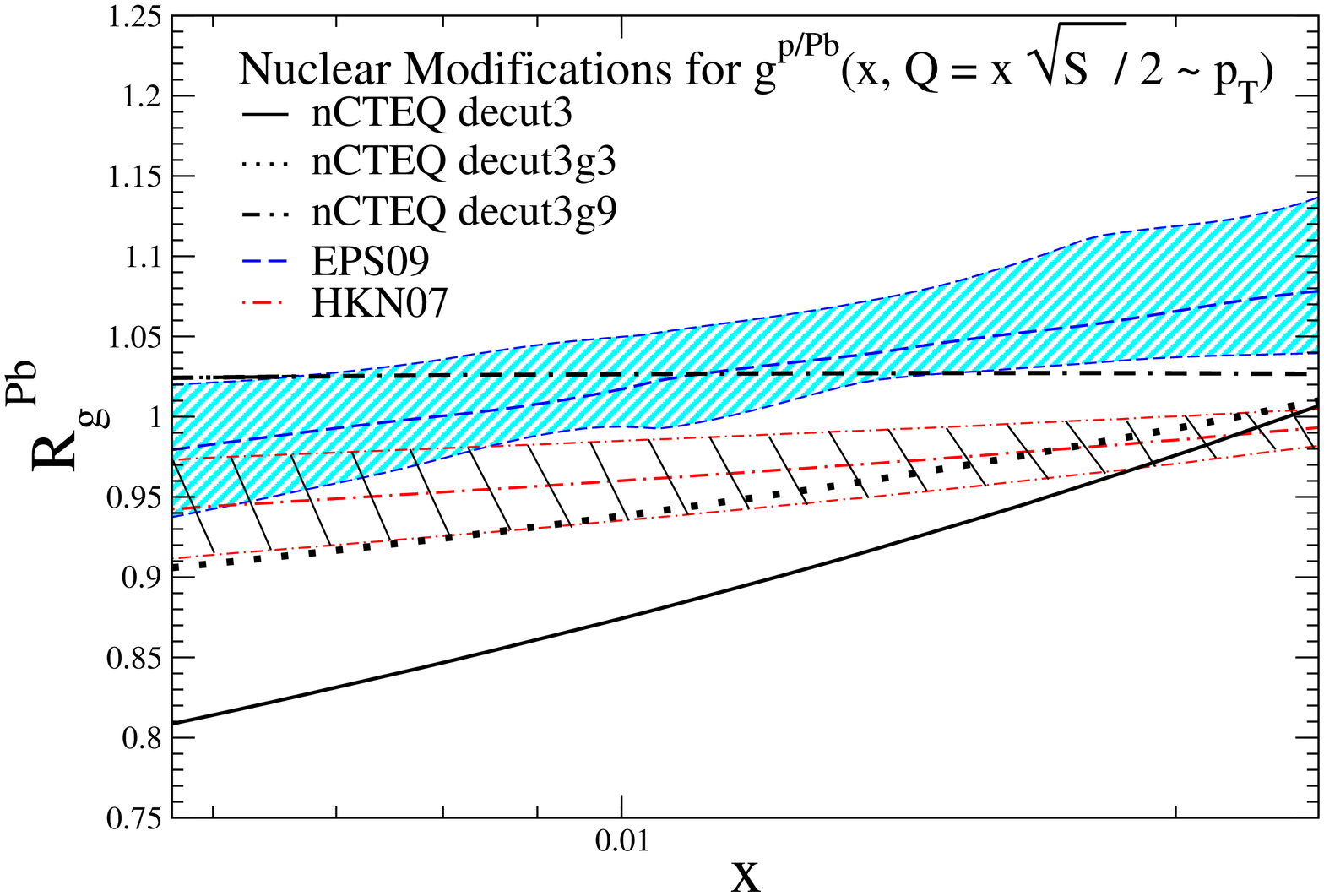}
\caption{Left: nuclear production ratio of $\gamma+c$ cross-section at LHC within ALICE PHOS acceptances, 
using nCTEQ decut3 (solid black line), nCTEQ decut3g3 (dotted black line), nCTEQ decut3g9 (dash-dot-dashed black line), EPS09 (dashed blue line) + error band, HKN07 (dash-dotted red line) + error band. 
Right: $R_g^{Pb}(x,Q=x\sqrt{S}/2 \sim p_T)$ ratio as a function of $x$, in the $x$ region probed at the LHC.
This figure corresponds to the enlargement of the box region in the right panel of Fig.~\protect\ref{fig:gluon}. 
}
\label{fig:ratio_alice}
\end{center}
\end{figure}

\subsection{Nuclear production ratios}

The nuclear production ratio $R_{pPb}^{\gamma+c}={1\over 208}
{d\sigma/dp_{T\gamma}(p{\rm Pb} \rightarrow \gamma+c+X)
\over d\sigma/dp_{T\gamma}(pp \rightarrow \gamma+c+X)}$
is shown in Fig.~\ref{fig:ratio_alice} (left) using the nCTEQ \texttt{decut3} (solid black line), 
nCTEQ \texttt{decut3g3} (dotted black line), nCTEQ \texttt{decut3g9} (dash-dot-dashed black line), 
EPS09 (dashed blue line), and HKN07 (dash-dotted red line) nuclear PDFs.
For the latter two cases the bands represent the nPDF uncertainties calculated as
described in section~\ref{sec:PDF}.  
Remarkably, there is almost no overlap between the EPS09 and the HKN predictions, therefore an 
appropriate measurement of this process will be able to distinguish 
between the two nPDF sets.  
The nCTEQ nuclear modification, using the \texttt{decut3} fit, is 
considerably different from the two other sets at lower values of $\pt$.
On the other hand, for the \texttt{decut3g9} set, the nuclear modification factor is close to unity in the
$x$-range as shown in Fig.~\ref{fig:ratio_alice} (right), giving rise to the nuclear production ratio for
this nCTEQ set which lies inside the EPS09 uncertainty band, Fig.~\ref{fig:ratio_alice} (left).
We stress again that both, \texttt{decut3} and \texttt{decut3g9}, are perfectly acceptable fits
to the $\ell A$ DIS+DY data with different assumptions on the small $x$ behavior.
We further show the ratio for \texttt{decut3g3}, as a representative lying between the two extremes.
Inspecting Fig.~\ref{fig:decut3gx2}, it is clear that the rest of the predictions from the \texttt{decut3g} series
would fill the gap between the \texttt{decut3} curve and the \texttt{decut3g9} curve.
Taken together, this gives a more realistic impression of the true PDF uncertainty of the nuclear 
production ratio. 
Therefore, measurements in this region will provide useful constraints on the nuclear gluon distribution.

Some further comments are in order: 
(i) In this paper we have demonstrated that
the ratio of the $\gamma + c$ cross-section in $pA$ over $pp$ collisions
at central rapidities will be very useful to constrain the nuclear gluon 
distribution.
At forward rapidities, even smaller $x_{_2}$ values could be probed in the nuclear targets where 
the uncertainties are largest.
At backward rapidities, large $x_{_2}$ is probed, hence the cross-section in this rapidity region will be sensitive to any existent intrinsic charm contribution in the nucleus.
Such a measurement could be performed with the CMS and ATLAS detectors which cover a wider range
in rapidity. We postpone such a study to a future publication, since currently there are no available IC nuclear PDFs; 
(ii) At the LHC $\gamma+b$ events will also be produced with sufficient statistics.
Experimentally this channel might be preferable due to the much better
$b$-tagging efficiencies.
Furthermore, uncertainties related to possible intrinsic charm contributions should be
much reduced in the bottom case.
However, as for $\gamma+c$ production,
the nuclear production ratios follow closely the gluon ratio and, 
therefore, we do not show a separate figure here.


\section{Conclusions}
\label{sec:conclusions}
We have performed a detailed phenomenological study of direct photon production
in association with a heavy-quark jet in $pA$ collisions at RHIC and at the LHC, at next-to-leading order in QCD.
The dominant contribution to this process is given by the 
$gQ\to \gamma Q[+g]$ subprocess. This offers a sensitive mechanism to constrain the heavy-quark and gluon 
distributions in nuclei, whose precise knowledge is necessary in order to predict the rates of hard processes 
in heavy-ion collisions where quark-gluon plasma is expected to be formed.

We have performed the calculation of $\gamma+Q$ production spectra at RHIC and at the LHC within 
the acceptances of various detectors (PHENIX and ALICE-PHOS/ALICE-EMCal) and have presented the corresponding counting rates. At the LHC the $\gamma+c$ and $\gamma+b$ production rate is important, while at RHIC only $\gamma+c$ events will be copiously produced.

Our results for RHIC (see Fig.~\ref{fig:ratio_rhic}) exhibit a strong sensitivity to the nuclear 
gluon distribution permitting to constrain it at 
$x \sim 0.1$--$0.2$.  
Similarly to RHIC the ratio at the LHC (see Fig.~\ref{fig:ratio_alice}) 
is very sensitive to the gluon distribution probing a smaller $x \sim 10^{-2}$, i.e. in a complementary range to RHIC.
These results have been obtained in the ``standard approach'' of radiatively 
generated charm distribution. 
A future study will focus on the possibility to constrain the intrinsic charm contribution to the 
nucleus as well as the proton.

 
\section*{Acknowledgment} 
We are much indebted to J.\ Faivre and C.\ Furget and G.\ Conesa for many helpful discussions
on the ALICE experiment in the initial stage of the project.
We are grateful to R. Granier de Cassagnac and H. Woehri for help with the experimental cuts 
for the CMS detector.
We also would like to thank Z. Conesa del Valle for useful remarks.
The work of T.~Stavreva was supported by a research grant of the University of Grenoble.
The work of F.~Arleo and I.~Schienbein was supported by a CNRS research 
grant ``PEPS Physique Th\'eorique aux Interfaces''.  This work was partially supported by the U.S.\ Department of Energy
under contract DE-FG02-04ER41299, and the Lightner-Sams Foundation.

\bibliographystyle{JHEP}
\bibliography{alice_TSv2} 

\end{document}